\documentclass{LMCS}

\def\dOi{10(1:13)2014}
\lmcsheading%
{\dOi}
{1--29}
{}
{}
{Dec.~\phantom01, 2011}
{Feb.~14, 2014}
{}

\ACMCCS{[{\bf Mathematics of computing}]: Probability and
  statistics---Stochastic processes---Markov processes; Design and
  analysis of algorithms---Mathematical optimization---Continuous
  optimization---Stochastic control and optimization\,/\,Convex
  Optimization; [{\bf Software and its engineering}]: Software
  creation and management---Software verification and
  validation---Formal software verification}

\def\techreport{}

\usepackage{mathtools, amsmath, amssymb}
\usepackage{ifthen}
\usepackage{color}
\usepackage{hyperref}
\usepackage{tikz}
\usepackage[format=hang]{subfig}
\usepgflibrary{shapes}
\usetikzlibrary{arrows,positioning}

\usepackage{flushend}

\tikzstyle{distr}=[inner sep=0mm, minimum size=1mm, draw, circle, fill]
\tikzstyle{state}=[inner sep=0.5mm, minimum size=3mm, draw]
\tikzstyle{trarr}=[semithick, -latex, rounded corners]

\newcommand{\WAIT}{\mathsf{Wait}}
\newcommand{\SWITCH}{\mathsf{Switch}}
\newcommand{\MECuni}{\mathcal{C}}
\newcommand{\MEC}{C}

\newcommand{\stabi}[1]{\kappa_{#1}}

\newcommand{\last}{\mathit{last}}

\newcommand{\dist}{\mathit{dist}}
\newcommand{\St}{\Sigma}

\newcommand{\StM}{\St^{M}}

\newcommand{\Prb}{\mathbb{P}}
\renewcommand{\Pr}[3]{\Prb^{#1}_{#2}\hspace{-0.16em}\left[{#3}\right]}   %
\newcommand{\Exp}{\mathbb{E}}
\newcommand{\Ex}[3]{\Exp^{#1}_{#2}\hspace{-0.16em}\left[{#3}\right]}   %
\newcommand{\reach}{\mathit{Reach}}
\newcommand{\lrLim}[1]{\mathrm{lr}(#1)}  %
\newcommand{\freq}[3]{\mathrm{freq}(#1,#2,#3)}  %

\newcommand{\tran}[1]{\stackrel{#1}\rightarrow}

\newcommand{\src}[1]{\mathit{Src}(#1)}
\newcommand{\act}[1]{\mathit{Act}(#1)}
\newcommand{\calF}{\mathcal{F}}

\newcommand{\Qset}{\mathbb{Q}}
\newcommand{\Nset}{\mathbb{N}}
\newcommand{\Rset}{\mathbb{R}}
\newcommand{\Zset}{\mathbb{Z}}

\begin{document}

\title[MDP with Multiple Long-run Average Objectives]{Markov Decision Processes with Multiple Long-run Average Objectives
}
\author[T.~Br\'azdil]{Tom\'a\v{s} Br\'{a}zdil\rsuper a}
\address{{\lsuper{a,b,e}}Faculty of Informatics\\
Masaryk University\\
Brno, Czech Republic}
\email{\{brazdil,xbrozek,kucera\}@fi.muni.cz}

\author[V.~Bro\v{z}ek]{V\'aclav~Bro\v{z}ek}
\address{\vspace{-18 pt}}

\author[K.~Chatterjee]{Krishnendu~Chatterjee\rsuper c}
\address{{\lsuper c}IST Austria\\
Klosterneuburg, Austria}
\email{krish.chat@ist.ac.at}

\author[V.~Forejt]{Vojt\v{e}ch~Forejt\rsuper d}
\address{{\lsuper d}Department of Computer Science\\
University of Oxford, UK}
\email{vojfor@comlab.ox.ac.uk}

\author[A.~Ku\v{c}era]{Anton\'in~Ku\v{c}era\rsuper e}
\address{\vspace{-18 pt}}

%

\begin{abstract}
  We study Markov decision processes (MDPs) with multiple
  limit-average (or mean-payoff) functions.  We consider two different
  objectives, namely, expectation and satisfaction objectives.  Given
  an MDP with $\pmb{k}$ limit-average functions, in the expectation objective the
  goal is to maximize the expected limit-average value, and in the satisfaction
  objective the goal is to maximize the probability of runs such that
  the limit-average value stays above a given vector.  
  We show that
  under the expectation objective, in contrast to the
  case of one limit-average function, both randomization and memory are necessary for strategies even for $\pmb{\varepsilon}$-approximation,
  and that finite-memory randomized strategies are sufficient for achieving Pareto optimal values.
  Under the satisfaction objective, in contrast to the
  case of one limit-average function, infinite memory is necessary 
  for strategies achieving a specific value (i.e. randomized finite-memory strategies are not sufficient), whereas memoryless randomized strategies 
  are sufficient for $\pmb{\varepsilon}$-approximation, for all 
  $\pmb{\varepsilon>0}$.  We further prove that
  the decision problems for both expectation and satisfaction
  objectives can be solved in polynomial time and the trade-off curve
  (Pareto curve) can be $\pmb{\varepsilon}$-approximated in time polynomial
  in the size of the MDP and $\pmb{\frac{1}{\varepsilon}}$, and exponential
  in the number of limit-average functions, for all $\pmb{\varepsilon>0}$.  Our
  analysis also reveals flaws in %
  previous work for MDPs with multiple
  mean-payoff functions under the expectation objective, corrects the
  flaws, and allows us to obtain improved results.
\end{abstract}

\maketitle\vfill

\section{Introduction}
\label{sec-intro}

Markov decision processes (MDPs) are the standard models 
for probabilistic dynamic systems that exhibit both probabilistic and
nondeterministic behaviors~\cite{Puterman,FV97}. 
In each state of an MDP, a controller chooses one of several actions 
(the nondeterministic choices), and the system stochastically evolves to a 
new state based on the current state and the chosen action.
A reward (or cost) is associated with each transition and the central 
question is to find a strategy of choosing the actions that optimizes 
the rewards obtained over the run of the system.
One classical way to combine the rewards over the run of the system
is the \emph{limit-average (or mean-payoff)} function that assigns to every
run the average of the rewards over the run.
MDPs with single mean-payoff functions have been widely studied in 
literature (see, e.g., \cite{Puterman,FV97}).
In many modeling domains, however, there is not a single goal to be 
optimized, but multiple, potentially dependent and conflicting goals.
For example, in designing a computer system, the goal is to maximize 
average performance while minimizing average power consumption.
Similarly, in an inventory management system, the goal is to optimize
several potentially dependent costs for maintaining each kind of product.
These motivate the study of MDPs with multiple mean-payoff functions.

Traditionally, MDPs with mean-payoff functions have been studied with only
the \emph{expectation} objective, where the goal is to maximize (or minimize) the expectation 
of the mean-payoff function. There are numerous applications of MDPs 
with expectation objectives in inventory control, planning, and 
performance evaluation~\cite{Puterman,FV97}. In this work we consider
both the expectation objective and also the 
\emph{satisfaction} objective for a given MDP. In both cases
we are given an MDP with $k$ reward functions, and the goal is 
to maximize (or minimize)
either the $k$-tuple of expectations, or 
the probability of runs such that the mean-payoff value stays above 
a given vector.

To get some intuition about the difference between the 
expectation/satisfaction objectives and to show that in
some scenarios the satisfaction objective is preferable,
consider a filehosting 
system
where the users can
download files at various speed, depending on the current setup and
the number of connected customers.
For simplicity, let us assume that
a user has 20\% chance to get a 2000kB/sec connection, and 80\%
chance to get a slow 20kB/sec connection. Then, the overall 
performance of the server can be reasonably measured by the 
expected amount of transferred data per user and second (i.e., the expected
mean payoff) which is 416kB/sec. However, a single user is
more interested in her  chance of downloading the files
quickly, which can be measured
by the probability of establishing and maintaining a reasonably 
fast connection \mbox{(say,~$\geq$\,1500kB/sec)}. Hence, the system 
administrator may want to maximize the expected mean payoff (by changing
the internal setup of the system), while a single user aims at maximizing 
the probability of satisfying her preferences (she can achieve that, e.g.,
by buying a priority access, waiting till 3 a.m., or simply connecting 
to a different server; obviously, she might also wish to minimize other 
mean payoffs such as the price per transferred bit). In other words, the 
expectation objective is relevant in situations when we are interested 
in the ``average'' behaviour of 
many instances of a given system, while the satisfaction objective 
is useful for analyzing and
optimizing particular executions.

In MDPs with multiple mean-payoff functions, various strategies may produce
incomparable solutions, and consequently there is no ``best'' solution
in general. Informally, the set of \emph{achievable solutions} 
\begin{enumerate}[label=(\roman*)]
\item under the  expectation  objective is the set of 
  all vectors $\vec{v}$ such that there is a strategy to ensure that 
  the expected mean-payoff value vector under the strategy is at 
  least $\vec{v}$; 
\item under the satisfaction objective is the set of tuples 
  $(\nu,\vec{v})$ where $\nu\in[0,1]$ and $\vec{v}$ is a vector such 
  that there is  a strategy under which with probability at least $\nu$
  the mean-payoff value vector of a run is at least~$\vec{v}$.
\end{enumerate}
The ``trade-offs'' among the goals represented by the individual
mean-payoff functions are formally captured by the \emph{Pareto curve},
which consists of all minimal tuples (wrt.\ componentwise ordering) 
that are not strictly dominated by any achievable solution. Intuitively, 
the Pareto curve consists of ``limits'' of achievable solutions, and in
principle it may contain tuples that are not achievable solutions
(see Section~\ref{sec-results}). Pareto optimality has been studied in 
cooperative game theory \cite{Owen95} and in multi-criterion 
optimization and decision making in both economics and engineering 
\cite{Koski,YC,Szymanek}.

Our study of MDPs with multiple mean-payoff functions is motivated
by the following fundamental questions, which concern both basic 
properties and algorithmic aspects of the expectation/satisfaction 
objectives:
\begin{enumerate}[label=Q.\arabic*]
\item What type of strategies is sufficient (and necessary) for
  achievable solutions?
\item Are the elements of the Pareto curve achievable solutions?
\item Is it decidable whether a given vector represents an
  achievable solution?
\item Given an achievable solution, is it possible to compute
  a strategy which achieves this solution? 
\item Is it decidable whether a given vector belongs to the
  Pareto curve?
\item Is it possible to compute a finite representation/approximation 
  of the Pareto curve?
\end{enumerate}
We provide comprehensive answers to the above questions, both
for the expectation and the satisfaction objective. We also analyze the
complexity of the problems given in Q.3--Q.6. From a practical point
of view, it is particularly encouraging that most of the considered
problems turn out to be solvable \emph{efficiently}, i.e., in polynomial time.
More concretely, our answers to Q.1--Q.6\ are the following:
\begin{enumerate}[label=4.b]
\item[1.a] For the expectation objectives, finite-memory randomized strategies are
  sufficient and necessary for all achievable solutions. Memory and randomization
  may also be needed to approximate an achievable solution up to $\varepsilon$ for
  a given $\varepsilon>0$.
\item[1.b] For the satisfaction objectives, achievable solutions require
  infinite memory in general, but memoryless randomized strategies are
  sufficient to approximate any achievable solution up to an arbitrarily
  small $\varepsilon >0$.
\item[2.] All elements of the Pareto curve are achievable solutions.
\item[3.] The problem  whether a given vector represents an
  achievable solution is solvable in polynomial time.
\item[4.a]  For the expectation objectives, a strategy which achieves
  a given solution is computable in polynomial time.   
\item[4.b]  For the satisfaction objectives, a strategy which 
  \mbox{$\varepsilon$-approximates} a given solution is computable in 
  polynomial time.
\item[5.] The problem  whether a given vector belongs to the Pareto curve
  is solvable in polynomial time.
\item[6.] A finite description of the Pareto curve is computable 
  in exponential time. Further, an  $\varepsilon$-approximate Pareto curve
  is computable in time which is polynomial in $1/\varepsilon$, the 
  size of a given MDP and the maximal absolute value of a reward assigned, and exponential in the number of 
  mean-payoff functions. 
\end{enumerate}
A more detailed and precise explanation of our results is postponed
to Section~\ref{sec-results}.

Let us note that MDPs with multiple mean-payoff functions under the 
expectation objective were also studied in~\cite{Cha07}, and it was 
claimed that memoryless randomized strategies are sufficient for
\mbox{$\varepsilon$-approximation} 
of the Pareto curve, for all $\varepsilon>0$, and an NP algorithm was 
presented to find a memoryless randomized strategy achieving a given vector. 
We show with an example that under the expectation objective 
there exists \mbox{$\varepsilon>0$} such that randomized 
strategies \emph{do require} 
memory for \mbox{$\varepsilon$-approximation}, and thus reveal 
a flaw in the earlier 
paper.

Similarly to the related papers~\cite{CMH06,EKVY08,FKN+11} (see Related Work),
we obtain our results by a characterization of the set of achievable 
solutions by a set of linear constraints, and from the linear constraints 
we construct witness strategies for any achievable solution.
However, our approach differs significantly from the previous work.
In all the previous works, the linear constraints are used to encode 
a \emph{memoryless} strategy either directly for the MDP~\cite{CMH06}, 
or (if memoryless strategies do not suffice in general) for a finite 
``product'' of the MDP and the specification function expressed as automata, from which
the memoryless strategy is then transferred to a finite-memory 
strategy for the original MDP~\cite{EKVY08,FKN+11,CY98}. 
In our setting new problems arise.
Under the expectation objective with mean-payoff function, 
neither is there any immediate notion of ``product'' of MDP and 
mean-payoff function and nor do memoryless strategies suffice.
Moreover, even for memoryless strategies the linear constraint 
characterization is not straightforward for mean-payoff functions, 
as in the case of discounted~\cite{CMH06}, reachability~\cite{EKVY08} 
and total reward functions~\cite{FKN+11}: for example, in~\cite{Cha07} 
even for memoryless strategies there was no linear constraint 
characterization for mean-payoff function and only an NP algorithm 
was given.
Our result, obtained by a characterization of linear constraints directly 
on the original MDP, requires involved and intricate construction 
of witness strategies. 
Moreover, our results are significant and non-trivial generalizations of the 
classical results for MDPs with a single mean-payoff function, where 
memoryless pure optimal strategies exist, while for multiple functions 
both randomization and memory is necessary.
Under the satisfaction objective, any finite product on which a memoryless strategy 
would exist is not feasible as in general witness strategies for achievable 
solutions may need an infinite amount of memory.
We establish a correspondence between the set of achievable solutions 
under both types of objectives for strongly connected MDPs. 
Finally, we use this correspondence to obtain our result for satisfaction objectives.

A conference version of this work was published at the conference LICS 2011 \cite{lics2011}.

\smallskip\noindent{\bf Related Work.} 
The study of Markov decision processes with multiple expectation objectives 
has been initiated in the area of applied probability theory, where it is 
known as {\em constrained MDPs}~\cite{Puterman, Altman}. 
The attention in the study of constrained MDPs has been focused mainly to 
restricted classes of MDPs, such as unichain MDPs where all states are 
visited infinitely often under any strategy. 
Such restriction both  guarantees the existence of
memoryless optimal strategies as well as simpler linear programming 
based algorithm for the problem, than the general case studied in this paper.

For general finite-state MDPs,~\cite{CMH06} studied MDPs with multiple discounted reward functions.
It was shown that memoryless strategies suffice for Pareto optimization,
and a polynomial-time algorithm was given to approximate
(up to a given relative error)
the Pareto curve by reduction to multi-objective linear programming 
and using the results of~\cite{PY00}.
MDPs with multiple qualitative $\omega$-regular specifications 
were studied in~\cite{EKVY08}.
It was shown that the Pareto curve can be approximated
in polynomial
time; the algorithm reduces the problem to MDPs with multiple
reachability specifications, which can be solved by multi-objective 
linear programming.
In~\cite{FKN+11}, the results of~\cite{EKVY08} were extended to combine
$\omega$-regular and expected total reward objectives.
MDPs with multiple mean-payoff functions under expectation objectives were 
considered in~\cite{Cha07}, and our analysis reveals flaws in the 
earlier paper, correct the flaws, and allows us to present significantly improved 
results (a polynomial-time algorithm for finding a strategy achieving
a given vector as compared to the previously suggested incorrect NP algorithm).
Moreover, the satisfaction objective has not been considered in
multi-objective setting before, and even in single objective case it has been
considered only in a very specific setting~\cite{BBE10}.

\newcommand{\pat}{\omega}
\newcommand{\Pat}{\mathsf{Runs}}
\newcommand{\fpat}{w}
\newcommand{\mem}{\mathsf{M}}
\newcommand{\Cone}{\mathsf{Cone}}
\newcommand{\reals}{\mathbb{R}}
\newcommand{\lrIf}[1]{\mathrm{lr}_{\mathrm{inf}}(#1)}  %
\newcommand{\lrSf}[1]{\mathrm{lr}_{\mathrm{sup}}(#1)}  %
\newcommand{\vare}{\varepsilon}
\newcommand{\AcEx}{\mathsf{AcEx}}
\newcommand{\AcPr}{\mathsf{AcSt}}

\section{Preliminaries}
We use  $\Nset$, $\Zset$, $\Qset$, and $\Rset$ 
to denote the sets of positive integers, integers, rational numbers, and real numbers, 
respectively. Given two vectors $\vec{v},\vec{u} \in \Rset^k$, where
$k \in \Nset$, we write $\vec{v} \leq \vec{u}$ iff 
$\vec{v}_i \leq \vec{u}_i$ for all $1 \leq i \leq k$, and
$\vec{v} < \vec{u}$ iff $\vec{v} \leq \vec{u}$ and 
$\vec{v}_i < \vec{u}_i$ for some $1 \leq i \leq k$. 

We assume familiarity with basic notions of probability
theory, e.g., \emph{probability space}, \emph{random variable}, or 
\emph{expected value}.
As usual, a \emph{probability distribution} over a finite or 
countably infinite set $X$ is a function
$f : X \rightarrow [0,1]$ such that \mbox{$\sum_{x \in X} f(x) = 1$}. 
We call $f$ \emph{positive} if 
$f(x) > 0$ for every $x \in X$, \emph{rational} if $f(x) \in
\Qset$ for every $x \in X$, and \emph{Dirac} if $f(x) = 1$ for some 
$x \in X$. The set of all distributions over $X$ is denoted by
$\dist(X)$.

\smallskip\noindent{\bf Markov chains.} 
A \emph{Markov chain} is a tuple \mbox{$M = (L,\tran{},\mu)$} where $L$ is
a finite or countably infinite set of locations, 
\mbox{${\tran{}} \subseteq L \times (0,1] \times L$} is a transition relation
such that for each fixed $\ell \in L$, $\sum_{\ell \tran{x} \ell'} x = 1$, and
$\mu$ is the initial probability distribution on~$L$.

A \emph{run} in $M$ is an infinite sequence $\pat = \ell_1 \ell_2 \ldots$
of locations such that $\ell_i \tran{x} \ell_{i{+}1}$ for every $i \in \Nset$.
A \emph{finite path} in $M$ is a finite prefix of a run. Each finite
path $\fpat$ in $M$ determines the set $\Cone(\fpat)$ consisting of
all runs that start with $\fpat$. To $M$ we associate the probability 
space $(\Pat_M,\calF,\mathbb{P})$, where $\Pat_M$ is the set of all 
runs in $M$, $\calF$ is the $\sigma$-field generated by all $\Cone(\fpat)$,
and $\mathbb{P}$ is the unique probability measure such that
$\mathbb{P}(\Cone(\ell_1,\ldots,\ell_k)) = 
\mu(\ell_1) \cdot \prod_{i=1}^{k-1} x_i$, where
$\ell_i \tran{x_i} \ell_{i+1}$ for all \mbox{$1 \leq i < k$} (the empty
product is equal to $1$).

\smallskip\noindent{\bf Markov decision processes.} 
A \emph{Markov decision process} (MDP) is a tuple of the form $G=(S,A,\mathit{Act},\delta)$ 
where $S$ is a \emph{finite} set of states, $A$ is a \emph{finite} set 
of actions, $\mathit{Act} : S\rightarrow 2^A\setminus \{\emptyset\}$ is 
an action enabledness 
function that assigns to each state $s$ the set $\act{s}$ of actions enabled 
at $s$, and $\delta : S\times A\rightarrow \dist(S)$ is a probabilistic 
transition function  that given a state $s$ and an action 
$a \in \act{s}$ enabled at $s$ gives a probability distribution over the 
successor states.
For simplicity, we assume that every action is enabled in exactly one state, 
and we denote this state $\src{a}$. Thus, henceforth we will assume that 
$\delta: A \rightarrow \dist(S)$.

A \emph{run} in $G$ is an infinite alternating sequence of states
and actions $\pat=s_1 a_1 s_2 a_2\ldots$
such that for all $i \geq 1$, $\src{a_i}=s_i$ and
$\delta(a_i)(s_{i+1}) > 0$. 
We denote by $\Pat_G$ the set of all runs in~$G$.
A \emph{finite path} of length~$k$ in~$G$ is a finite prefix
$\fpat = s_1 a_1\ldots a_{k-1} s_k$ of a run in~$G$.
For a finite path $\fpat$ we denote by $\last(\fpat)$ the last 
state of~$w$.

A pair $(T,B)$ with $\emptyset\neq T\subseteq S$ and $B\subseteq \bigcup_{t\in T}\act{t}$
is an \emph{end component} of $G$
if (1) for all $a\in B$, whenever $\delta(a)(s')>0$ then $s'\in T$;
and (2) for all $s,t\in T$ there is a finite path 
$\pat = s_1 a_1\ldots a_{k-1} s_k$ such that $s_1 = s$, $s_k=t$, and all states
and actions that appear in $\pat$ belong to $T$ and $B$, respectively.
An end component $(T,B)$ is a \emph{maximal end component (MEC)}
if it is maximal wrt.\ pointwise subset ordering. Given an end component
$C=(T,B)$, we sometimes abuse notation by using $C$ instead of $T$ or $B$, 
e.g.,
by writing $a\in C$ instead of $a\in B$ for $a\in A$.

\smallskip\noindent{\bf Strategies and plays.} 
Intuitively, a strategy in an MDP $G$ is a ``recipe'' to choose actions.
Usually, a strategy is formally defined as a function 
$\sigma : (SA)^*S \to \dist(A)$ that given a finite path~$\fpat$, representing 
the history of a play, gives a probability distribution over the 
actions enabled in~$\last(\fpat)$. In this paper, we adopt a somewhat
different
\ifthenelse{\isundefined{\techreport}}{%
(though equivalent---see~\cite{techreport})}{%
(though equivalent -- see Section~\ref{app-strat-eq})}
definition, which allows a more natural
classification of various strategy types. Let $\mem$ be a finite or countably
infinite set of \emph{memory elements}. A \emph{strategy}
is a triple
$\sigma = (\sigma_u,\sigma_n,\alpha)$, where 
$\sigma_u: A\times S \times \mem \to \dist(\mem)$ and 
$\sigma_n: S \times \mem \to \dist(A)$ are \emph{memory update}
and \emph{next move} functions, respectively, and $\alpha$ is
an initial distribution on memory elements. We require that for 
all $(s,m) \in S \times \mem$, the distribution $\sigma_n(s,m)$ assigns a
positive value only to actions enabled at~$s$. The set of all
strategies is denoted by $\St$ (the underlying MDP~$G$ will be always
clear from the context).

Let $s \in S$ be an initial state. A \emph{play} of $G$ determined by 
$s$ and a strategy $\sigma$ is a Markov chain 
$G^\sigma_s$ (or just $G^\sigma$ if $s$ is clear from the context)
where the set of locations is $S \times \mem \times A$,
the initial distribution $\mu$ is positive only on (some) elements 
of $\{s\} \times  \mem \times A$ where
$\mu(s,m,a) = \alpha(m) \cdot \sigma_n(s,m)(a)$, 
and $(t,m,a) \tran{x} (t',m',a')$ iff 
\[ 
   x ~~ = ~~
   \delta(a)(t')  \cdot \sigma_u(a,t',m)(m') \cdot \sigma_n(t',m')(a')
     ~~ > ~~  0
\;.
\]
Hence, $G^\sigma_s$ starts in a location chosen randomly according
to $\alpha$ and $\sigma_n$. In a current location $(t,m,a)$, 
the next action to be performed is $a$, hence the probability of entering
$t'$ is $\delta(a)(t')$. The probability of updating the memory to $m'$
is $\sigma_u(a,t',m)(m')$, and the probability of selecting $a'$ as 
the next action is $\sigma_n(t',m')(a')$. We assume that these choices
are independent, and thus obtain the product above. 

In this paper, we consider various functions over $\Pat_G$ that 
become random variables over $\Pat_{G^\sigma_s}$ after fixing 
some $\sigma$ and~$s$. For example, for $F \subseteq S$
we denote by $\reach(F)\subseteq\Pat_G$ the set of all runs
reaching $F$.
Then  $\reach(F)$ naturally determines
$\reach^\sigma_s(F) \subseteq \Pat_{G^\sigma_s}$
by simply ``ignoring'' the visited memory
elements. To simplify and unify our notation, we write, e.g.,
$\Pr{\sigma}{s}{\reach(F)}$ instead of 
  $\Pr{\sigma}{s}{\reach^\sigma_s(F)}$, where
  $\mathbb{P}^\sigma_s$ is the probability measure of the probability
  space associated to $G^\sigma_s$.
We also adopt this notation for other events and functions, such as
$\lrIf{\vec{r}}$ or $\lrSf{\vec{r}}$ defined in the next section, and write, e.g.,
$\Ex{\sigma}{s}{\lrIf{\vec{r}}}$ instead of 
$\Ex{}{}{\lrIf{\vec{r}}^\sigma_s}$.

\smallskip\noindent{\bf Strategy types.}
In general, a strategy may use infinite memory, and both 
$\sigma_u$ and $\sigma_n$ may randomize. According to the use of 
randomization, a strategy, $\sigma$, can be classified
as 
\begin{itemize}
\item \emph{pure} (or \emph{deterministic}), if $\alpha$ is Dirac and
  both the memory update and the next move function give a Dirac
  distribution for every argument;
\item \emph{deterministic-update}, if $\alpha$ is Dirac and 
  the memory update function gives a Dirac distribution for
  every argument;
\item \emph{stochastic-update}, if $\alpha$, $\sigma_u$, and $\sigma_n$
  are unrestricted.
\end{itemize}
Note that every pure strategy is deterministic-update, and every
deterministic-update strategy is stochastic-update. 
A \emph{randomized} strategy is a strategy which is not necessarily pure.
We also classify the strategies according to the size of memory
they use. Important subclasses are 
\emph{memoryless} strategies, in which $\mem$ is a singleton,
\emph{$n$-memory} strategies, in which $\mem$ has exactly $n$~elements, and
\emph{finite-memory} strategies, in which $\mem$ is finite.
By $\StM$ we denote the set of all memoryless strategies.
Memoryless strategies can be specified as 
$\sigma: S {\to} \dist(A)$.
\emph{Memoryless pure} strategies, i.e., those which are 
both pure and memoryless, can be specified as $\sigma: S {\to} A$.

For a finite-memory strategy $\sigma$, a 
\emph{bottom strongly connected component} (BSCC) of $G^\sigma_s$ is 
a subset of locations \mbox{$W \subseteq S\times \mem \times A$} 
such that for all $\ell_1 \in W$ and \mbox{$\ell_2 \in S\times \mem\times A$}
we have that (i) if $\ell_2$ is reachable from $\ell_1$, then 
$\ell_2 \in W$, and (ii) for all $\ell_1,\ell_2 \in W$ we have that
$\ell_2$ is reachable from $\ell_1$.
Every BSCC $W$ determines a unique end component 
$(\{s\mid (s,m,a)\in W\},\{a\mid (s,m,a)\in W\})$
of $G$, and we sometimes do not strictly distinguish between $W$ 
and its associated end component.

As we already noted, stochastic-update strategies can be easily
translated into ``ordinary'' strategies of the form 
\mbox{$\sigma : (SA)^*S \to \dist(A)$}, and vice versa
\ifthenelse{\isundefined{\techreport}}{%
(see \cite{techreport}).
}{%
(see Section~\ref{app-strat-eq}).
}
Note that a finite-memory stochastic-update strategy $\sigma$ can be easily
implemented by a \emph{stochastic finite-state automaton} that scans
the history of a play ``on the fly'' (in fact, $G^\sigma_s$ simulates 
this automaton). Hence, finite-memory stochastic-update strategies
can be seen as natural extensions of ordinary (i.e., 
deterministic-update) finite-memory strategies that are 
implemented by deterministic finite-state automata.

\begin{figure}
\begin{tikzpicture}[every node/.style={inner sep=0.5mm}][font=\small]
\node[state] at (0,0) (s1) {$s_1$};
\node[distr] at (1,0) (s1a1) {};
\node[distr] at (0,-0.5) (s1a2) {};
\node[state] at (2,0) (s2) {$s_2$};
\node[distr] at (2.5,0) (s2a3) {};
\node[state] at (0,-1.5) (s3) {$s_3$};
\node[distr] at (1,-1.5) (s3a4) {};
\node[distr] at (-0.5,-1.5) (s3a5) {};
\node[state] at (2,-1.5) (s4) {$s_4$};
\node[distr] at (1,-2) (s4a6) {};

\draw (0.15,-0.5) arc (0:-90:0.15);
\draw (1.15,-1.5) arc (0:90:0.15);
\draw[trarr] (s1) -- (s1a1) node[midway,above] {$a_1$} -- (s2);
\draw[trarr] (s1) -- (s1a2) node[midway,left] {$a_2$} -- (s3) node[pos=0.2,left] {$0.5$};
\draw[trarr] (s1a2) -| (s2) node[pos=0.25,above] {$0.5$};
\draw[trarr] (s3) -- (s3a4) node[midway,above] {$a_4$} -- (s4) node[midway,above] {$0.3$};
\draw[trarr] (s3a4) -- +(0,0.5) -| (s3.50)  node[pos=0.2,above] {$0.7$};
\draw[trarr] (s4) |- (s4a6) node[pos=0.75,above] {$a_6$} -| (s3);

\draw[trarr] (s3) -- +(-0.5,-0.5) -- (s3a5) node[midway,left] {$a_5$}-- +(0,0.5) -- (s3);
\draw[trarr] (s2) -- +(0.5,-0.5) -- (s2a3) node[midway,right] {$a_3$} -- +(0,0.3) -- (s2);
\end{tikzpicture}
\hspace*{1.5cm}
\begin{tikzpicture}[every node/.style={inner sep=0.5mm}][font=\small]

\node[state] at (0,0) (s1m1a1) {$(s_1,m_1,a_2)$};
\node[state] at (3,0) (s1m1a2) {$(s_1,m_1,a_1)$};
\node[state] at (0,-0.9) (s3m1a5) {$(s_3,m_1,a_5)$};
\node[state] at (3,-0.9) (s2m1a3) {$(s_2,m_1,a_3)$};
\node[state] at (0,-1.8) (s3m2a4) {$(s_3,m_2,a_4)$};
\node[state] at (3,-1.8) (s4m2a6) {$(s_4,m_2,a_6)$};

\draw[trarr] (-1.5,0) -- (s1m1a1) node[pos=0.4,below] {$0.5$};
\draw[trarr] (4.5,0) -- (s1m1a2) node[pos=0.4,below] {$0.5$};
\draw[trarr] (s1m1a1) -- (s3m1a5) node[midway,left] {$0.5$};
\draw[trarr] (s1m1a1) -- (s2m1a3) node[midway,above] {$0.5$};
\draw[trarr] (s1m1a2) -- (s2m1a3) node[midway,right] {$1$};
\draw[trarr] (s3m1a5.west) -- +(-0.5,-0.4) -- +(-0.5,0.4) node[midway,left] {$0.5$} -- (s3m1a5.west);
\draw[trarr] (s2m1a3.east) -- +( 0.5,-0.4) -- +( 0.5,0.4) node[midway,right] {$1$} -- (s2m1a3.east);
\draw[trarr] (s3m1a5) -- (s3m2a4) node[midway,left] {$0.5$};
\draw[trarr] (s3m2a4.10) -- (s4m2a6.170) node[midway,above] {$0.3$};
\draw[trarr] (s3m2a4.west) -- +(-0.5,-0.4) -- +(-0.5,0.4) node[midway,left] {$0.7$} -- (s3m2a4.west);
\draw[trarr] (s4m2a6.-170) -- (s3m2a4.-10) node[midway,below] {$1$};
\end{tikzpicture}
\caption{Running example MDP (left) and its play (right)\label{fig:running-both}}
\end{figure}
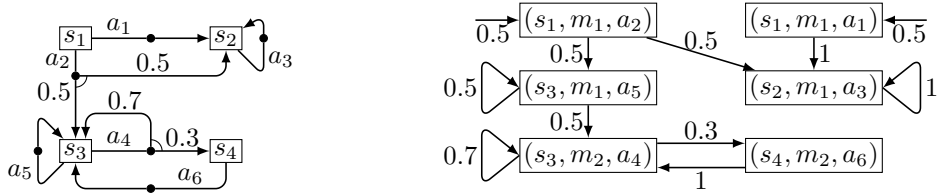

\smallskip\noindent{\bf A running example (I).}
  As an example, consider the MDP $G=(S,A,\mathit{Act},\delta)$
  of~Figure~\ref{fig:running-both}~(left).  Here, $S=\{s_1,\ldots,s_4\}$,
  $A=\{a_1,\ldots,a_6\}$, $\mathit{Act}$ is denoted using the
  labels on lines going from actions,
  e.g., $\mathit{Act}(s_1)=\{a_1,a_2\}$, and $\delta$ is given by the
  arrows, e.g., $\delta(a_4)(s_4)=0.3$. Note that $G$ has four  end
  components (one on $\{s_2\}$, another on $\{s_3\}$, and two on $\{s_3,s_4\}$) and two MECs.

  Let $s_1$ be the initial state and $\mem=\{m_1,m_2\}$. Consider a
  stochastic-update finite-memory strategy
  $\sigma=(\sigma_u,\sigma_n,\alpha)$ where $\alpha$ chooses $m_1$
  deterministically, and $\sigma_n(m_1,s_1)=[a_1\mapsto 0.5,
  a_2\mapsto 0.5]$, $\sigma_n(m_2,s_3)=[a_4\mapsto 1]$ and otherwise
  $\sigma_n$ chooses self-loops. The memory update function $\sigma_u$
  leaves the memory intact except for the case $\sigma_u(m_1,s_3)$
  where both $m_1$ and $m_2$ are chosen with probability $0.5$. The
  play $G^\sigma_{s_1}$ is depicted in Figure~\ref{fig:running-both}~(right).
\section{Main Results}
\label{sec-results}

In this paper we establish basic results about Markov decision
processes with \emph{expectation} and \emph{satisfaction} objectives
specified by multiple \emph{limit-average} (or \emph{mean-payoff})
functions. We adopt the variant where rewards 
are assigned to edges (i.e., actions) rather than states of a given MDP.

Let $G=(S,A,\mathit{Act},\delta)$ be a MDP, and $r : A \to
\Qset$ a \emph{reward function}.  Note that $r$ may also take negative
values. For every $j \in \Nset$, let $A_j: \Pat_G \to A$ be a function which to
every run $\pat \in \Pat_G$ assigns the $j$-th action of $\pat$. Since
the limit-average function $\lrLim{r} : \Pat_G \rightarrow \Rset$
given by
\[
  \lrLim{r}(\pat)  =  \lim_{T\rightarrow\infty} 
     \frac{1}{T}\sum_{t=1}^{T} {r(A_t(\pat))} 
\]
may be undefined for some runs, we consider its lower and upper 
approximation $\lrIf{r}$ and $\lrSf{r}$ that are defined for
\emph{all} $\pat \in \Pat$ as follows: 
\begin{eqnarray*}
  \lrIf{r}(\pat) & = & \liminf_{T\rightarrow\infty} 
     \frac{1}{T}\sum_{t=1}^{T} {r(A_t(\pat))},\\
  \lrSf{r}(\pat) & = & \limsup_{T\rightarrow\infty} 
     \frac{1}{T}\sum_{t=1}^{T} {r(A_t(\pat))}
.
\end{eqnarray*}
For a vector $\vec{r} = (r_1,\ldots,r_k)$ of reward functions, we 
similarly define the $\Rset^k$-valued functions 
\begin{eqnarray*}
 \lrLim{\vec{r}} & = & (\lrLim{r_1},\ldots,\lrLim{r_k}),\\
 \lrIf{\vec{r}}  & = & (\lrIf{r_1},\ldots,\lrIf{r_k}),\\
 \lrSf{\vec{r}}  & = & (\lrSf{r_1},\ldots,\lrSf{r_k}).
\end{eqnarray*}
We sometimes refer to ``runs satisfying $\lrLim{\vec{r}} \ge \vec{v}$''
instead of ``runs $\omega$ satisfying $\lrLim{\vec{r}}(\omega) \ge \vec{v}$''.

Now we introduce the expectation and satisfaction objectives
determined by~$\vec{r}$. 
\begin{itemize}
\item The \emph{expectation} objective amounts to maximizing or minimizing 
  the expected value of $\lrLim{\vec{r}}$. Since $\lrLim{\vec{r}}$ may
  be undefined for some runs, we actually aim at maximizing the 
  expected value of $\lrIf{\vec{r}}$ or minimizing the expected value of 
  $\lrSf{\vec{r}}$ (wrt.\ componentwise ordering $\leq$). 
\item The \emph{satisfaction} objective means maximizing the probability
  of all runs where $\lrLim{\vec{r}}$ stays above or below a given vector 
  $\vec{v}$. Technically, we aim at maximizing the probability
  of all runs where $\lrIf{\vec{r}} \geq \vec{v}$, or at maximizing the probability of all runs where
  $\lrSf{\vec{r}} \leq \vec{v}$. 
\end{itemize}
The expectation objective is relevant in situations when we are interested 
in the average or aggregate behaviour of many instances of a system, and 
in contrast, the satisfaction objective is relevant when we are interested 
in particular executions of a system and wish to optimize the probability 
of generating the desired executions.
Since $\lrIf{\vec{r}} = -\lrSf{-\vec{r}}$, the problems of 
maximizing and minimizing the expected value of 
$\lrIf{\vec{r}}$ and $\lrSf{\vec{r}}$ are dual. Therefore, we 
consider just the problem of maximizing the expected value 
of $\lrIf{\vec{r}}$. For the same reason, we consider only 
the problem of maximizing the probability
of all runs where $\lrIf{\vec{r}} \geq \vec{v}$. 

If $k$ (the dimension of $\vec{r}$) is at least two, there
might be several incomparable solutions to the expectation objective;
and if $\vec{v}$ is slightly changed, the achievable 
probability of all runs satisfying $\lrIf{\vec{r}} \geq \vec{v}$ may 
change considerably. Therefore, we aim not only at constructing 
a particular solution, but on characterizing and approximating the 
whole space of \emph{achievable solutions} for the expectation/satisfaction 
objective. Let $s \in S$ be some (initial) state of $G$. We define the sets  
$\AcEx(\lrIf{\vec{r}})$ and $\AcPr(\lrIf{\vec{r}})$ of \emph{achievable
vectors} for the expectation and satisfaction objectives as follows:
\begin{eqnarray*}
 \AcEx(\lrIf{\vec{r}})  & \!\!\!\!\!  =  & \!\!\!\!\!
   \{\vec{v} \mid \exists \sigma \in \Sigma: 
     \ \Ex{\sigma}{s}{\lrIf{\vec{r}}} \geq \vec{v} \},\\
 \AcPr(\lrIf{\vec{r}}) & \!\!\!\!\!  =  & \!\!\!\!\! 
   \{(\nu, \vec{v}) \mid \exists \sigma \in \Sigma:
     \ \Pr{\sigma}{s}{\lrIf{\vec{r}}\geq\vec{v}} \; \geq \; \nu \}.
\end{eqnarray*}%
Intuitively, if $\vec{v},\vec{u}$ are achievable vectors such that
$\vec{v} > \vec{u}$, then $\vec{v}$ represents a ``strictly better''
solution than $\vec{u}$. The set of ``optimal'' solutions defines
the \emph{Pareto curve} for $\AcEx(\lrIf{\vec{r}})$ and 
$\AcPr(\lrIf{\vec{r}})$. In general, the Pareto curve 
for a given set $Q \subseteq \reals^k$ is the set $P$ of all minimal
vectors $\vec{v} \in \Rset^k$ such $\vec{v} \not< \vec{u}$ for all
$\vec{u} \in Q$. Note that $P$ may contain vectors that are not
in~$Q$ (for example, if  $Q = \{x \in \Rset \mid x<2\}$, then
$P = \{2\}$). However, every vector $\vec{v} \in P$ is ``almost'' in~$Q$
in the sense that for every $\vare > 0$ there is $\vec{u} \in Q$ with
$\vec{v} \leq \vec{u} + \vec{\vare}$, where 
$\vec{\vare} = (\vare,\dots,\vare)$. 
This naturally leads
to the notion of an \emph{\mbox{$\vare$-approximate} Pareto curve}, $P_\vare$,
which is a subset of $Q$ such that for all vectors $\vec{v} \in P$ 
of the Pareto curve there is a vector $\vec{u} \in P_\vare$ such that 
$\vec{v} \leq \vec{u} + \vec{\vare}$.
Note that $P_{\vare}$ is not unique.

\smallskip\noindent{\bf A running example (II).}
Consider again the MDP~$G$ of Figure~\ref{fig:running-both}~(left), and the strategy
$\sigma$ constructed in our running example~(I).
Let $\vec{r} = (r_1,r_2)$, where
$r_1(a_6)=1$, $r_2(a_3)=2$, $r_2(a_4)=1$, 
and otherwise the rewards are zero. Let 
\[
  \omega=(s1,m1,a_2)(s_3,m_1,a_5)\big((s_3,m_2,a_4)(s_4,m_2,a_6)\big)^\omega
\]
Then $\lrLim{\vec{r}}(\omega)=(0.5,0.5)$. Considering the 
expectation objective, we have that 
$\Ex{\sigma}{s_1}{\lrIf{\vec{r}}}  = 
(\frac{3}{52}, \frac{22}{13})$. Considering the satisfaction objective, 
we have that $(0.5,0,2)\in\AcPr(\vec{r})$ because 
$\Pr{\sigma}{s_1}{\lrIf{\vec{r}}\geq (0,2)} = 0.5$.
The Pareto curve for $\AcEx(\lrIf{\vec{r}})$ consists of the points
$
 \{(\frac{3}{13} x, \frac{10}{13}x + 2(1{-}x))
  \mid 0\le x \le 0.5\}
$,
and the Pareto curve for $\AcPr(\lrIf{\vec{r}})$
is $\{(1,0,2)\}\cup \{(0.5,x,1-x) \mid 0 < x_1\le \frac{10}{13}\}$.
\smallskip

Now we are equipped with all the notions needed for understanding the main 
results of this paper. Our work is motivated by the six fundamental 
questions given in Section~\ref{sec-intro}. In the next subsections
we give detailed answers to these questions.

\subsection{Expectation objectives}
\label{sec-exp-obj}

The answers to Q.1-Q.6\ for  the expectation objectives
are the following: 

\begin{enumerate}[label=A.\arabic*]
\item
   For all achievable solutions, $2$-memory stochastic-update strategies are sufficient, i.e., for all $\vec{v} \in 
   \AcEx(\lrIf{\vec{r}})$ there is a $2$-memory stochastic-update 
   strategy $\sigma$ satisfying $\Ex{\sigma}{s}{\lrIf{\vec{r}}} \geq \vec{v}$.
\item
   The Pareto curve~$P$ for $\AcEx(\lrIf{\vec{r}})$ is a subset
   of $\AcEx(\lrIf{\vec{r}})$, i.e., all optimal solutions are
   achievable. 
\item
   There is a polynomial-time algorithm which, given any
   \mbox{$\vec{v} \in \Qset^k$}, decides
   whether $\vec{v} \in\AcEx(\lrIf{\vec{r}})$.
\item
   If $\vec{v} \in \AcEx(\lrIf{\vec{r}})$, then there is 
   a $2$-memory stochastic-update strategy $\sigma$ constructible
   in polynomial time 
   satisfying $\Ex{\sigma}{s}{\lrIf{\vec{r}}} \geq \vec{v}$.
\item
   There is a polynomial-time algorithm which, given 
   \mbox{$\vec{v} \in \mathbb{R}^k$}, decides
   whether $\vec{v}$ belongs to the Pareto curve for 
   $\AcEx(\lrIf{\vec{r}})$.
\item
  There is a convex hull $Z$ of finitely many vectors such that:
  $\AcEx(\lrIf{\vec{r}})$ is a downward closure of $Z$ (i.e. $\AcEx(\lrIf{\vec{r}}) = \{\vec{v}\mid \exists \vec{u} \in Z : \vec{v}\le \vec{u}\}$);
   The Pareto curve for $\AcEx(\lrIf{\vec{r}})$ is a union of all
   facets of $Z$ whose vectors are not strictly dominated by
   vectors of $Z$.
   Further, an $\vare$-approximate Pareto curve for $\AcEx(\lrIf{\vec{r}})$
   is computable in time polynomial in  $\frac{1}{\vare}$, $|G|$, 
   and $\max_{a\in A}\max_{1\le i\le k} |\vec{r}_i(a)|$, and exponential in~$k$.
\end{enumerate}
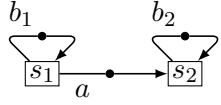
\begin{figure}
{\centering \begin{tikzpicture}

\node[state] at (0,0) (s0) {$s_1$};
\node[distr] at (0,0.5) (s0a) {};
\node[distr] at (0.9,0) (s0b) {};
\node[state] at (1.9,0) (s1) {$s_2$};
\node[distr] at (1.9,0.5) (s1c) {};

\draw[trarr] (s0) -- (s0b) node[midway,below] {$a$} -- (s1);
\draw[trarr] (s0) -- +(-0.5,0.5) -- (s0a) node[pos=0.5,above] {$b_1$} -- +(0.5,0) -- (s0);
\draw[trarr] (s1) -- +(-0.5,0.5) -- (s1c) node[pos=0.5,above] {$b_2$} -- +(0.5,0) -- (s1);
\end{tikzpicture}}
\caption{Example of insufficiency of memoryless strategies\label{fig:nonmem}}
\end{figure}
Let us note that A.1~is tight in the sense that neither memoryless randomized 
nor pure strategies are sufficient for achievable solutions. This is
witnessed by the MDP of Figure~\ref{fig:nonmem} with reward functions
$r_1$, $r_2$ such that $r_i(b_i) = 1$ and $r_i(b_j) = 0$ for $i\neq j$.
Consider a strategy $\sigma$ which initially selects between the actions
$b_1$ and $a$ randomly (with probability~$0.5$) and
then keeps selecting $b_1$ or $b_2$, whichever is available. Hence,
$\Ex{\sigma}{s_1}{\lrIf{(r_1,r_2)}} = (0.5,0.5)$. However,
the vector $(0.5,0.5)$ is not achievable by a strategy~$\sigma'$
which is memoryless or pure, because then we inevitably have that
$\Ex{\sigma'}{s_1}{\lrIf{(r_1,r_2)}}$ is equal either to
$(0,1)$ or $(1,0)$. The example also shows that memory and randomization is
needed for $\varepsilon$-approximation. Considering e.g. $\varepsilon=0.1$,
a history-dependent randomized strategy is needed to achieve the value
$(0.5-0.1,0.5-0.1)$ or better.

The $2$-memory stochastic-update strategy from A.1~and A.4~operates in two modes.
Starting in the first mode, it reaches the MECs of the MDP with appropriate probabilities;
once a MEC is reached, the strategy stochastically switches to a second mode, never leaving
the current MEC and ensuring certain ``frequencies'' of taking the actions of the MEC.
Since both modes can be implemented by memoryless strategies, we get that we only require two memory
elements to remember which mode is currently being executed.
We also show that the $2$-memory stochastic-update strategy constructed
can be efficiently transformed into
a finite-memory deterministic-update randomized strategy, and hence the answers
A.1 and A.4 are also valid for \emph{finite-memory 
deterministic-update randomized} strategies
(see Section~\ref{app-expsem-det-up}).
Observe that A.2 can be seen as a generalization of the well-known result
for single payoff functions which says that finite-state MDPs with
mean-payoff objectives have optimal strategies (in this case, the
Pareto curve consists of a single number known as the ``value''). Also
observe that A.2 does \emph{not} hold for infinite-state MDPs
(a counterexample is simple to construct even for a single reachability objective, see
e.g.~\cite[Example 6]{IC08}).

Finally, note that if $\sigma$ is a finite-memory stochastic-update 
strategy, then $G_s^\sigma$ is a \emph{finite-state} Markov chain.
Hence, for almost all runs $\pat$ in $G_s^\sigma$ we have that
$\lrLim{\vec{r}}(\pat)$ exists and it is equal to 
$\lrIf{\vec{r}}(\pat)$. This means that there is actually no
difference between maximizing the expected value of
$\lrIf{\vec{r}}$ and maximizing the expected value of $\lrLim{\vec{r}}$
over all strategies for which $\lrLim{\vec{r}}$ exists.

\subsection{Satisfaction objectives}
\label{sec-results-sat}

The answers to Q.1-Q.6\ for the satisfaction objectives 
are presented below.

\begin{enumerate}[label=B.\arabic*]
\item Achievable vectors require strategies with
  infinite memory in general.
  However, memoryless randomized strategies are sufficient for 
  $\vare$-approximate achievable vectors; in fact, a stronger claim holds and
  for every $\vare>0$ and 
  $(\nu,\vec{v}) \in \AcPr(\lrIf{\vec{r}})$, there is a memoryless
  randomized strategy $\sigma$ with
  \[
   \Pr{\sigma}{s}{\lrIf{\vec{r}} \geq\vec{v} - \vec{\vare}} 
   \; \geq \; \nu.
  \] 
  Here $\vec{\vare} = (\vare,\ldots,\vare)$.
\item
  The Pareto curve~$P$ for $\AcPr(\lrIf{\vec{r}})$ is a subset
  of $\AcPr(\lrIf{\vec{r}})$, i.e., all optimal solutions are
  achievable. 
\item
  There is a polynomial-time algorithm which, given $\nu \in [0,1]$ and
  $\vec{v} \in \Qset^k$, decides
  whether $(\nu,\vec{v}) \in \AcPr(\lrIf{\vec{r}})$. 
\item
  If $(\nu,\vec{v}) \in \AcPr(\lrIf{\vec{r}})$,
  then for every $\vare>0$ there is a memoryless randomized strategy
  $\sigma$ constructible in polynomial time such that
  $\Pr{\sigma}{s}{\lrIf{\vec{r}} \geq\vec{v} - \vec{\vare}} 
  \; \geq \; \nu-\vare$. 
\item
   There is a polynomial-time algorithm which, given \mbox{$\nu \in [0,1]$} and
   $\vec{v} \in \mathbb{R}^k$, decides
   whether $(\nu,\vec{v})$ belongs to the Pareto curve for 
   $\AcPr(\lrIf{\vec{r}})$.
\item
   The Pareto curve $P$ for $\AcPr(\lrIf{\vec{r}})$ may be neither connected, nor closed.
   However, $P$ is a union of finitely many
   sets whose closures are convex polytopes, and, perhaps surprisingly, the set $\{\nu \mid (\nu,\vec{v})\in P\}$
   is always finite. The sets in the union that gives $P$ (resp.\ the inequalities that
   define them) can be computed.
   Further, an $\vare$-approximate Pareto curve for $\AcPr(\lrIf{\vec{r}})$
   is computable in time polynomial in $\frac{1}{\vare}$, $|G|$, 
   and $\max_{a\in A}\max_{1\le i\le k} |\vec{r}_i(a)|$, and exponential in~$k$.
\end{enumerate}
The algorithms of B.3 and B.4 are polynomial in the size of $G$ and the size 
of binary representations of $\vec{v}$ and $\frac{1}{\vare}$.

The result B.1 is again tight.
In Lemma~\ref{lem-eps-mp} we show that
memoryless pure strategies
are insufficient for $\vare$-approximate achievable vectors,
i.e.,
there are $\vare>0$ and $(\nu,\vec{v}) \in \AcPr(\lrIf{\vec{r}})$ 
such that for every memoryless pure
strategy $\sigma$ we have
$\Pr{\sigma}{s}{\lrIf{\vec{r}}\geq\vec{v} - \vec{\vare}} < \nu-\vare$.

As noted in B.1, a strategy $\sigma$ achieving a given vector 
$(\nu,\vec{v}) \in \AcPr(\lrIf{\vec{r}})$ may require infinite memory. 
Still, our proof of B.1 reveals a ``recipe'' for 
constructing such a $\sigma$ by simulating the  memoryless randomized 
strategies $\sigma_\varepsilon$ which $\vare$-approximate $(\nu,\vec{v})$ 
(intuitively, for smaller and 
smaller~$\varepsilon$, the strategy $\sigma$ simulates 
$\sigma_\varepsilon$ longer and longer; the details are discussed 
in Section~\ref{sec-probsem}). Hence, for almost all runs 
$\pat$ in $G_s^\sigma$ we again have that
$\lrLim{\vec{r}}(\pat)$ exists and it is equal to 
$\lrIf{\vec{r}}(\pat)$.

\section{Solution for Expectation Objectives}\label{sect-expsem}
The technical core of our results for expectation objectives 
is the following:
\begin{figure}[t]\small
\begin{align}
\mathbf{1}_{s_0}(s) + \sum_{a\in A} y_{a}\cdot \delta(a)(s) & = 
 \sum_{a\in \act{s}} y_{a} + y_s &\text{for all  $s\in S$}
\label{eq:ya}\\
\sum_{s\in S_{\textit{MEC}}}y_{s} & =  1 &
\label{eq:ys1}\\
 \sum_{s\in C} y_{s} & =  \sum_{a\in A\cap C} x_{a} &
   \text{for all MEC $C$ of $G$}
\label{eq:yC}\\
\sum_{a\in A} x_{a}\cdot \delta(a)(s) & = 
\sum_{a\in \act{s}} x_{a} & \text{for all  $s\in S$}
\label{eq:xa}\\
\sum_{a\in A} x_{a}\cdot\vec{r}_i(a) & \ge  \vec{v}_i &
\text{ for all $1\le i \le k$}
\label{eq:rew}
\end{align}
\caption{System $L$ of linear inequalities for Theorem~\ref{thrm-exp}. (We define
$S_{\textit{MEC}}\subseteq S$ to be the states contained in some MEC of $G$, 
$\mathbf{1}_{s_0}(s)=1$ 
if $s = s_0$, and $\mathbf{1}_{s_0}(s)=0$ otherwise.)}
\label{system-L}
\end{figure}
\begin{thm}
\label{thrm-exp}
Let $G=(S,A,\mathit{Act},\delta)$ be an MDP, $s_0 \in S$ an initial state, $\vec{r} = (r_1,\ldots,r_k)$
a tuple of reward functions, and $\vec{v} \in \Rset^k$.
The system of linear inequalities~$L$ from Figure~\ref{system-L}
is constructible in polynomial time and satisfies:
\begin{itemize}
\item\label{thrm-exp-1} every nonnegative solution of $L$ induces a 
  $2$-memory stochastic-update
  strategy~$\sigma$ satisfying $\Ex\sigma{s_0}{\lrIf{\vec{r}}}\geq \vec{v}$;
\item\label{thrm-exp-2} if $\vec{v} \in  \AcEx(\lrIf{\vec{r}})$, then 
  $L$ has a nonnegative solution.
\end{itemize}
\end{thm}

As we already noted in Section~\ref{sec-intro}, the proof of 
Theorem~\ref{thrm-exp} is non-trivial and it is based on novel techniques 
and observations. Our results about expectation objectives are
corollaries to Theorem~\ref{thrm-exp} and the arguments developed in its 
proof. For the rest of this section, we fix an MDP $G$, 
a vector of rewards, $\vec{r}=(r_1,\ldots,r_k)$,
and an initial state $s_0$ (in the considered plays of $G$, the initial
state is not written explicitly, unless it is different from $s_0$).

Obviously, $L$ is constructible in polynomial time.  
Let us briefly explain the intuition behind $L$.
As mentioned earlier,
a $2$-memory stochastic-update
strategy witnessing that $\vec{v} \in  \AcEx(\lrIf{\vec{r}})$ works in two modes.
In the first mode it ensures that each MEC is reached and never left
with certain probability, and in the second mode actions are taken with
required frequencies. In $L$, the probability of reaching a MEC $C$
is encoded as the value $\sum_{s\in C} y_s$, and Equations~(\ref{eq:ya})
are used to ensure that the numbers obtained are indeed realisable under some strategy.
The meaning of these equations is similar as the meaning of similar equations in~\cite{EKVY08},
essentially the equations encode that the expected number of times a state is entered (left-hand side of the equations)
is equal to the expected number of times a state is left together with probability of switching to the second mode
(right-hand side of the equations). A more formal explanation of these equations is given at the end
of the proof of Proposition~\ref{prop:strat-sol}.
The frequency of taking an action $a$ is then encoded as $x_a$, and realisability
of the solution by some strategy is ensured using Equations~(\ref{eq:xa}). Here the meaning of the equations is
that the frequency with which a state is entered must be equal to the frequency with which it is left;
this is formalised in~Lemma~\ref{lem:xsol}.

As both directions of
Theorem~\ref{thrm-exp} are technically involved, we prove them
separately as
Propositions~\ref{prop:sol-strat}~and~\ref{prop:strat-sol}.
\begin{prop}\label{prop:sol-strat}
  Every nonnegative solution of the system $L$ of Figure~\ref{system-L} induces a $2$-memory
  stochastic-update strategy~$\sigma$ satisfying
  $\Ex\sigma{s_0}{\lrIf{\vec{r}}}\geq \vec{v}$.
\end{prop}

\subsubsection*{Proof of Proposition~\ref{prop:sol-strat}}
First, let us consider Equations~(\ref{eq:xa}) of~$L$. Intuitively, this
equation is solved by an ``invariant'' distribution on actions, i.e., each 
solution gives frequencies of actions (up to a multiplicative constant) 
defined for all $a\in A$, $s\in S$, and $\sigma \in \Sigma$ by
\[
\freq{\sigma}{s}{a}
\coloneqq
\lim_{T\rightarrow \infty} \frac{1}{T}\sum_{t=1}^{T} \Pr{\sigma}{s}{A_t=a}
,
\]
assuming that the defining limit exists (which might not be the
case---cf.\ the proof of Proposition~\ref{prop:strat-sol}).
We prove the following:
\begin{lem}\label{lem:xsol}
Assume that assigning (nonnegative) values
$\bar{x}_a$ to $x_a$ solves Equations~(\ref{eq:xa}).
Then there is a memoryless strategy $\xi$ such that for every
BSCCs $D$ of $G^{\xi}$, every $s\in D\cap S$, and every 
$a\in D\cap A$, we have that
$\freq{\xi}{s}{a}$ equals a common value
$\freq{\xi}{D}{a} \coloneqq \bar{x}_a/\sum_{a'\in D\cap A} \bar{x}_{a'}$.
\end{lem}
\begin{proof}
For all $s\in S$ we set $\bar{x}_s=\sum_{b\in\act{s}}\bar{x}_b$ and
define $\xi$ by
$
\xi(s)(a)
\coloneqq
\frac{\bar{x}_a}{\bar{x}_s}
$
if $\bar{x}_s>0$, and arbitrarily otherwise.
We claim that the vector of values $\bar{x}_s$ forms an invariant measure
of $G^{\xi}$. Indeed,
noting that
$
\sum_{a\in \act{s}} \xi(s)(a)\cdot \delta(a)(s')
$
is the probability of the transition $s\tran{}s'$ in $G^{\xi}$:
\begin{align*}
\sum_{s\in S} \bar{x}_s\cdot \sum_{a\in \act{s}} \xi(s)(a)\cdot \delta(a)(s')
&=
\sum_{s\in S} \sum_{a\in \act{s}} \bar{x}_s \cdot \frac{\bar{x}_a}{\bar{x}_s}\cdot \delta(a)(s')\\
&= \sum_{a\in A} \bar{x}_a \cdot \delta(a)(s')\\
&= \sum_{a\in \act{s'}} \bar{x}_a\tag{By Equation~\ref{eq:xa}}\\
&= \bar{x}_{s'} 
.
\end{align*}
As a consequence, $\bar{x}_s>0$ iff $s$ lies in some BSCC of $G^\xi$.
Choose some BSCC $D$, and denote by $\bar{x}_D$
the number
$
\sum_{a\in D\cap A} \bar{x}_{a}
=
\sum_{s\in D\cap S} \bar{x}_{s}
.
$
Also denote by $I_t^a$ the indicator of $A_t=a$, given by
$I_t^a=1$ if $A_t=a$ and $0$ otherwise.
By the Ergodic theorem for finite Markov chains (see, e.g.~\cite[Theorem~1.10.2]{Norris98}), for all
$s\in D\cap S$ and $a \in D \cap A$ we have
\[
\Ex\xi{s}{\lim_{T\to \infty} \frac{1}{T}\sum_{t=1}^{T} I_t^a}
=
\sum_{s'\in D\cap S}
\frac{\bar{x}_{s'}}{\bar{x}_D}\cdot \xi(s')(a)
=
\frac{\bar{x}_{s'}}{\bar{x}_D}\cdot\frac{\bar{x}_a}{\bar{x}_{s'}}
=
\frac{\bar{x}_{a}}{\bar{x}_D}
.
\]
Because $|I_t^a|\leq1$, Lebesgue Dominated convergence theorem
(see, e.g.~\cite[Chapter~4, Section~4]{Royden88})
yields
$
\Ex\xi{s}{\lim_{T\to \infty} \frac{1}{T}\sum_{t=1}^{T} I_t^a}
=
\lim_{T\to \infty} \frac{1}{T}\sum_{t=1}^{T} \Ex\xi{s}{I_t^a}
$
and thus
$
\freq{\xi}{s}{a}
=
\frac{\bar{x}_{a}}{\bar{x}_D}
=
\freq{\xi}{D}{a}
.
$
This finishes the proof of Lemma~\ref{lem:xsol}.
\end{proof}

Assume that the system $L$ is solved by
assigning nonnegative values $\bar{x}_a$ to $x_a$ and $\bar{y}_{\chi}$
to $y_{\chi}$ where $\chi\in A\cup S$. W.l.o.g. assume that $\bar{y}_s = 0$ for
all states $s$ not contained in any MEC. Let $\xi$ be the strategy of
Lemma~\ref{lem:xsol}. Using Equations~(\ref{eq:ya}), (\ref{eq:ys1}),
and (\ref{eq:yC}), we will define a 2-memory stochastic update strategy
$\sigma$ as follows.  The strategy $\sigma$ has two memory elements,
$m_1$ and $m_2$. A run of $G^{\sigma}$ starts in $s_0$ with a given
distribution on memory elements (see below). Then $\sigma$
plays according to a suitable memoryless strategy (constructed below)
until the memory changes to $m_2$, and then it starts behaving as $\xi$
forever. Given a BSCC $D$ of $G^{\xi}$, we denote by
$\Pr\sigma{s_0}{\text{switch to $\xi$ in $D$}}$ the probability that
$\sigma$ switches from $m_1$ to $m_2$ while in $D$.  We construct $\sigma$
so that
\begin{equation}\label{eq:switch}
\Pr\sigma{s_0}{\text{switch to $\xi$ in $D$}}\quad = \quad \sum_{a\in D\cap A} \bar{x}_a
\;.
\end{equation}
Then for all $a\in D\cap A$ we have
$\freq{\sigma}{s_0}{a}
=
\Pr\sigma{s_0}{\text{switch to $\xi$ in $D$}}\cdot \freq{\xi}{D}{a}
=
\bar{x}_a$.
Finally, we obtain the following:
\begin{equation}\label{eq:lrinf-freq}
\Ex\sigma{s_0}{\lrIf{\vec{r}_i}} = \sum_{a\in A} \vec{r}_i(a)\cdot \bar{x}_a 
\;.
\end{equation}
The equation can be derived as follows:
\begin{align*}
\Ex\sigma{s_0}{\lrIf{r_i}}
& =  \Ex\sigma{s_0}{\liminf_{T\to \infty} \frac{1}{T}\sum_{t=1}^{T}r_i(A_t)} \tag{definition} \\
& =  \Ex\sigma{s_0}{\lim_{T\to \infty} \frac{1}{T}\sum_{t=1}^{T}r_i(A_t)} \tag{see below} \\
& =  \lim_{T\to \infty} \frac{1}{T}\sum_{t=1}^{T} \Ex\sigma{s_0}{r_i(A_t)} \tag{see below}\\
& =  \lim_{T\to \infty} \frac{1}{T}\sum_{t=1}^{T} \sum_{a\in A} r_i(a)\cdot  \Pr{\sigma}{s_0}{A_t=a} \tag{definition of expectation} \\
& =  \sum_{a\in A} r_i(a)\cdot\lim_{T\to \infty} \frac{1}{T}\sum_{t=1}^{T} \Pr{\sigma}{s_0}{A_t=a} \tag{linearity of the limit}\\
& =  \sum_{a\in A} r_i(a)\cdot \freq{\sigma}{s_0}{a} \tag{definition of $\freq{\sigma}{s_0}{a}$} \\
& =  \sum_{a\in A} r_i(a)\cdot \bar{x}_a \tag{$\freq{\sigma}{s_0}{a}=\bar{x}_a$}
.
\end{align*}
The second equality follows from the fact that the limit is almost surely defined,
following from the Ergodic theorem applied to the BSCCs of the finite Markov chain $G^{\sigma}$.
The third equality holds by Lebesgue Dominated convergence theorem, because
$
\left|r_i(A_t)\right|
\leq
\max_{a\in A} |r_i(a)|
.
$
Note that the right-hand side of Equation~(\ref{eq:lrinf-freq})
is greater than or equal to $\vec{v}_i$ by Inequality~(\ref{eq:rew}) of~$L$.

So, it remains to construct the strategy $\sigma$ with the desired
``switching'' property expressed by Equations~(\ref{eq:switch}). 
Roughly speaking, we proceed in two steps. 
\begin{enumerate}[label=\arabic*.]
\item  We construct a \emph{finite-memory}
  stochastic update strategy $\bar{\sigma}$ satisfying 
  Equations~(\ref{eq:switch}). 
  The strategy $\bar{\sigma}$ is constructed so that it 
  initially behaves as a certain finite-memory stochastic update 
  strategy, but eventually this mode is ``switched'' to the strategy
  $\xi$ which is followed forever.
\item The only problem with $\bar{\sigma}$ is that it may use
  more than two memory elements in general. This is solved by applying
  the results of~\cite{EKVY08} and reducing the ``initial part''
  of $\bar{\sigma}$ (i.e., the part before the switch) into 
  a \emph{memoryless} strategy. Thus, we transform $\bar{\sigma}$
  into an ``equivalent'' strategy $\sigma$ which is 2-memory
  stochastic update.
\end{enumerate}
Now we elaborate the two steps.
 
\textit{Step 1.}
For every MEC $C$ of $G$, we denote by $y_C$ the
number $\sum_{s\in C} \bar{y}_s=\sum_{a\in A\cap C} \bar{x}_a$.  By
combining the solution of $L$ with the results of Sections~3 and 5
of~\cite{EKVY08}
one can construct a finite-memory
stochastic-update strategy $\zeta$ which stays eventually in each MEC
$C$ with probability $y_C$. Formally, the construction is
captured in the following lemma.
\begin{lem}
\label{lem-reach-MEC}
Consider numbers $\bar{y}_\chi$ for all $\chi\in S\cup A$
such that the assignment $y_\chi \coloneqq \bar{y}_\chi$
is a part of some nonnegative solution to $L$.
Then there is a finite-memory stochastic update
strategy $\zeta$
which, starting from $s_0$, stays eventually in each MEC $C$ with probability
$
y_C \coloneqq
\sum_{s\in C} \bar{y}_s
.
$
\end{lem}

\begin{proof}
In order to be able to use results of~\cite[Section~3]{EKVY08} we modify
the MDP $G$ and obtain a new MDP $G'$ as follows:
For each state $s$ we add a new absorbing state, $d_s$.
The only available action for $d_s$ leads to a loop transition back to $d_s$
with probability $1$. We also add a new action, $a_s^d$, to every $s\in S$.
The distribution associated with $a_s^d$ assigns probability $1$ to $d_s$.

Let us call $K$ the set of constraints of the LP on Figure~3 in \cite{EKVY08}.
From the values $\bar{y}_\chi$ we now construct a solution to $K$:
for every state $s\in S$ and every action $a\in\act{s}$
we set $y_{(s,a)}\coloneqq \bar{y}_a$,
and $y_{(s,a_s^d)}\coloneqq \bar{y}_s$.
The values of the rest of variables in $K$ are determined by the second set
of equations in $K$. The nonnegative constraints in $K$ are satisfied since
$\bar{y}_\chi$ are nonnegative.
Finally, the equations (\ref{eq:ya}) from $L$ imply that the first set of equations
in $K$ are satisfied, because $\bar{y}_\chi$ are part of a solution to $L$.

By Theorem~3.2 of~\cite{EKVY08} we thus have a memoryless strategy $\varrho$ for $G'$
which satisfies
$
\Pr{s_0}\varrho{\text{$\reach(d_s)$}}\geq y_s
$
for all $s\in S$.
The strategy $\zeta$ then mimics the behavior of $\varrho$ until the moment
when $\varrho$ chooses an action to enter some of the new absorbing states.
From that point on, $\zeta$ may choose some arbitrary fixed behavior to
stay in the current MEC (note that if the current state $s$ is not included in any MEC, then $\bar{y}_s=0$
and so the strategy $\varrho$ would not choose to enter the new absorbing state).
As a consequence:
$
\Pr{s_0}\zeta{\text{stay eventually in $C$}}\geq y_C
,
$
and in fact, we get equality here, because of the equations (\ref{eq:ys1}) from $L$.
Note that $\zeta$ only needs a finite constant amount of memory.
\end{proof}

The strategy $\bar\sigma$ works as follows. For a run initiated in $s_0$,
the strategy $\bar\sigma$ plays according to $\zeta$ until a
 BSCC of $G^{\zeta}$ is reached. This means that every possible 
 continuation of the path stays in the current MEC $C$ of $G$.
Assume that $C$ has states $s_1,\ldots,s_k$.  We denote by $\bar{x}_s$ the sum
$\sum_{a\in\act{s}}\bar{x}_a$.  At this point, the strategy
$\bar\sigma$ changes its behavior as follows: First, the strategy
$\bar\sigma$ strives to reach $s_1$ with probability one. Upon
reaching $s_1$, it chooses (randomly, with probability
$\frac{\bar{x}_{s_1}}{y_C}$) either to behave as $\xi$ forever, or to
follow on to $s_2$.
If the strategy $\bar\sigma$ chooses to go on to $s_2$, it strives 
to reach $s_2$ with probability one. Upon reaching $s_2$,
the strategy $\bar\sigma$ chooses (randomly, with probability 
$\frac{\bar{x}_{s_2}}{y_C-\bar{x}_{s_1}}$) 
either to behave as $\xi$ forever, or to follow on to $s_3$, and so, till 
$s_k$. That is, the probability of switching to $\xi$ in $s_i$ is 
$\frac{\bar{x}_{s_i}}{y_C-\sum_{j=1}^{i-1}\bar{x}_{s_j}}$.

Since $\zeta$ stays in a MEC $C$ with probability $y_C$, the
probability that the strategy $\bar\sigma$ switches to $\xi$ 
in $s_i$ is equal to $\bar{x}_{s_i}$. However, then for every BSCC $D$
of $G^{\xi}$ satisfying $D\cap C \neq\emptyset$ (and thus $D\subseteq
C$) we have that the strategy $\bar\sigma$ switches to $\xi$ in a
state of $D$ with probability $\sum_{s\in D\cap S}
\bar{x}_{s}=\sum_{a\in D\cap A} \bar{x}_{a}$. Hence, $\bar{\sigma}$
satisfies Equations~(\ref{eq:switch}).

\textit{Step 2.}  Now we show how to reduce the first phase of
$\bar{\sigma}$ (before the switch to $\xi$) into a memoryless
strategy, using the results of \cite[Section~3]{EKVY08}.
Unfortunately, these results are not applicable directly. 
We need to modify the MDP $G$ into a new MDP
$G'$, same as we did above: For each state $s$ we add a new absorbing state,
$d_s$.  The only available action for $d_s$ leads to a loop transition
back to $d_s$ with probability $1$. We also add a new action, $a_s^d$,
to every $s\in S$.  The distribution associated with $a_s^d$ assigns
probability~$1$ to~$d_s$.

Let us consider a finite-memory stochastic-update strategy, $\sigma'$,
for $G'$ defined as follows. The strategy $\sigma'$ behaves as
$\bar{\sigma}$ before the switch to $\xi$. Once $\bar{\sigma}$
switches to $\xi$, say in a state $s$ of $G$ with probability $p_s$,
the strategy $\sigma'$ chooses the action $a_s^d$ with probability
$p_s$. It follows that the probability of $\bar{\sigma}$ switching in
$s$ is equal to the probability of reaching $d_s$ in $G'$ under
$\sigma'$.  By \cite[Theorem~3.2]{EKVY08}, there is a memoryless
strategy, $\sigma''$, for $G'$ that reaches $d_s$ with probability
$p_s$. We define $\sigma$ in $G$ to behave as $\sigma''$ with the
exception that, in every state $s$, instead of choosing an action
$a_s^d$ with probability $p_s$ it switches to behave as $\xi$ with
probability $p_s$ (which also means that the initial distribution on
memory elements assigns $p_{s_0}$ to $m_2$).
Then, clearly, $\sigma$ satisfies Equations~(\ref{eq:switch}) because
\[
\Pr\sigma{s_0}{\text{switch in $D$}} = \sum_{s\in D}\Pr{\sigma''}{s_0}{\text{fire $a_s^d$}} = \sum_{s\in D}\Pr{\sigma'}{s_0}{\text{fire $a_s^d$}}
  = \Pr{\bar{\sigma}}{s_0}{\text{switch in $D$}} = \sum_{a\in D\cap A} \bar{x}_{a}.
\]
This concludes the proof of Proposition~\ref{prop:sol-strat}. \hfill $\Box$
\smallskip

\begin{prop}\label{prop:strat-sol}
If $\vec{v} \in  \AcEx(\lrIf{\vec{r}})$, then $L$ has a nonnegative solution.
\end{prop}
\begin{proof}
Let $\varrho\in\St$ be a strategy such that
$
\Ex\varrho{s_0}{\lrIf{\vec{r}}}
\geq
\vec{v}
$.
In general, the frequencies $\freq\varrho{s_0}{a}$ of the actions may 
not be well defined, because the defining limits may not exist. A crucial
trick to overcome this difficulty is to pick suitable ``related'' values, 
$f(a)$, lying between
$
\liminf_{T\rightarrow \infty} \frac{1}{T}\sum_{t=1}^{T} \Pr{\varrho}{s_0}{A_t=a}
$
and
$
\limsup_{T\rightarrow \infty} \frac{1}{T}\sum_{t=1}^{T} \Pr{\varrho}{s_0}{A_t=a}
$, which can be safely substituted for $x_a$ in~$L$. Since every infinite
sequence contains an infinite convergent subsequence, there is an increasing
sequence of indices, $T_0, T_1, \ldots$, such that the following limit exists
for each action $a\in A$
\[
f(a)
\coloneqq
\lim_{\ell\to\infty}
\frac{1}{T_\ell}
\sum_{t=1}^{T_\ell} \Pr\varrho{s_0}{A_t=a} \ .
\]
Setting
$
x_a \coloneqq f(a)
$
for all $a\in A$
satisfies Inequalities~(\ref{eq:rew}) and Equations~(\ref{eq:xa}) of~$L$.
Indeed, the former follows from
\mbox{$
\Ex\varrho{s_0}{\lrIf{\vec{r}}}
\geq
\vec{v}
$}
and the following inequality, which holds for all $1\leq i \leq k$:
\begin{equation}\label{eq:freq-lrinf}\small
\sum_{a\in A} \vec{r}_i(a) \cdot f(a) \ \geq \ \Ex\varrho{s_0}{\lrIf{\vec{r}_i}}
.
\end{equation}
The inequality follows from the following derivation:
\begin{align*}
\sum_{a\in A} r_i(a) \cdot f(a)
& = 
\sum_{a \in A}r_i(a)\cdot
\lim_{\ell\to\infty}
\frac{1}{T_\ell}
\sum_{t=1}^{T_\ell} \Pr{\varrho}{s_0}{A_t=a}\tag{definition of $f(a)$}\\
& = 
\lim_{\ell\to\infty}
\frac{1}{T_\ell}
\sum_{t=1}^{T_\ell} \sum_{a\in A}r_i(a)\cdot\Pr{\varrho}{s_0}{A_t=a}\tag{linearity of the limit}\\
& \geq 
\liminf_{T\to\infty}
\frac{1}{T}
\sum_{t=1}^T \sum_{a\in A}r_i(a)\cdot\Pr{\varrho}{s_0}{A_t=a}\tag{definition of $\liminf$}\\
& \geq 
\liminf_{T\to\infty}
\frac{1}{T}
\sum_{t=1}^T \Ex{\varrho}{s_0}{r_i(A_t)}\tag{linearity of the expectation}\\
& \geq
\Ex\varrho{s_0}{\lrIf{r_i}}\tag{see below}
.
\end{align*}
The last inequality is a consequence of
Fatou's lemma (see, e.g.~\cite[Chapter~4, Section~3]{Royden88})
-- although the function $r_i(A_t)$ may not be nonnegative, we can replace
it with the nonnegative function $r_i(A_t)-\min_{a\in A}r_i(a)$ and add the subtracted
constant afterwards.

To prove that Equations~(\ref{eq:xa}) are satisfied, it suffices to show that
for all $s\in S$ we have 
\begin{equation}\label{eq:invf}\small
\sum_{a\in A} f(a)\cdot \delta(a)(s) = \sum_{a \in \act{s}} f(a)
.
\end{equation}
This holds, because
\begin{align*}
\sum_{a\in A} f(a)\cdot \delta(a)(s) & =
  \sum_{a\in A}\lim_{\ell\to\infty} \frac{1}{T_\ell} \sum_{t=1}^{T_\ell} \Pr\varrho{s_0}{A_t=a}
   \cdot \delta(a)(s)\tag{definition of $f$}\\
   & = \lim_{\ell\to\infty} \frac{1}{T_\ell} \sum_{t=1}^{T_\ell}
   \sum_{a\in A} \Pr\varrho{s_0}{A_t=a}
   \cdot \delta(a)(s) \tag{linearity of the limit}\\
   & = 
   \lim_{\ell\to\infty} \frac{1}{T_\ell} \sum_{t=1}^{T_\ell}
   \Pr\varrho{s_0}{S_{t+1}=s} \tag{definition of $\delta$}\\
   & = 
   \lim_{\ell\to\infty} \frac{1}{T_\ell} \sum_{t=1}^{T_\ell}
   \Pr\varrho{s_0}{S_{t}=s}\tag{see below} \\
   & = 
   \lim_{\ell\to\infty} \frac{1}{T_\ell} \sum_{t=1}^{T_\ell}
   \sum_{a \in \act{s}}
   \Pr\varrho{s_0}{A_{t}=a} \tag{$s$ must be followed by $a\in \act{s}$}\\
   & = 
   \sum_{a \in \act{s}}
   \lim_{\ell\to\infty} \frac{1}{T_\ell} \sum_{t=1}^{T_\ell}
   \Pr\varrho{s_0}{A_{t}=a} \tag{linearity of the limit}\\
   & =  
   \sum_{a \in \act{s}}
    f(a)\tag{definition of $f$}
    \;.
\end{align*}
The fourth equality follows from the following:
\begin{align*}
\lim_{\ell\to\infty} \frac{1}{T_\ell} \sum_{t=1}^{T_\ell}
\Pr\varrho{s_0}{S_{t+1}=s}
-
\lim_{\ell\to\infty} \frac{1}{T_\ell} \sum_{t=1}^{T_\ell}
\Pr\varrho{s_0}{S_{t}=s}
&=
\lim_{\ell\to\infty} \frac{1}{T_\ell} \sum_{t=1}^{T_\ell}
(
\Pr\varrho{s_0}{S_{t+1}=s}
-
\Pr\varrho{s_0}{S_{t}=s}
)
\\
&=
\lim_{\ell\to\infty} \frac{1}{T_\ell}
(
\Pr\varrho{s_0}{S_{T_\ell+1}=s}
-
\Pr\varrho{s_0}{S_{1}=s}
)\\
&=0
.
\end{align*}

Now we have to set the values for $y_\chi$, $\chi\in A\cup S$,
and prove that they satisfy the rest of $L$ when the values $f(a)$ are 
assigned to $x_a$. Note that almost every run of $G^{\varrho}$ eventually 
stays in some MEC of $G$ (cf., e.g., \cite[Proposition~3.1]{CY98}).
For every MEC $C$ of $G$, let $y_C$ be the probability of all runs
in $G^{\varrho}$ that eventually stay in~$C$. Note that
\begin{equation}\small
\label{eq:fa-yc}
\begin{split}
\sum_{a\in A\cap C}
f(a)
& =
\sum_{a\in A\cap C}
\lim_{\ell\to\infty} 
\frac{1}{T_\ell}
\sum_{t=1}^{T_\ell} \Pr\varrho{s_0}{A_t=a} \\
& =
\lim_{\ell\to\infty} 
\frac{1}{T_\ell}
\sum_{t=1}^{T_\ell}
\sum_{a\in A\cap C}
\Pr\varrho{s_0}{A_t=a} \\
& =
\lim_{\ell\to\infty}
\frac{1}{T_\ell}
\sum_{t=1}^{T_\ell} \Pr\varrho{s_0}{A_t\in C} 
 = y_C .
\end{split}
\end{equation}
Here the last equality follows from the fact that 
$\lim_{\ell\to\infty} \Pr\varrho{s_0}{A_{T_{\ell}}\in C}$ 
is equal to the probability of all runs in $G^{\varrho}$ that 
eventually stay in~$C$ 
(recall that almost every run stays eventually in a MEC of $G$) and
the fact that the Ces\`{a}ro sum of a convergent sequence is equal 
to the limit of the sequence.

To obtain $y_a$ and $y_s$, we need to simplify the behavior of 
$\varrho$ before reaching a MEC for which we use
the results of~\cite{EKVY08}. As in the proof of 
Proposition~\ref{prop:sol-strat}, we first need to modify the
MDP $G$ into another MDP $G'$ as follows:
For each state $s$ we add a new absorbing state, $d_s$.
The only available action for $d_s$ leads to a loop transition back to $d_s$
with probability $1$. We also add a new action, $a_s^d$, to every $s\in S$.
The distribution associated with $a_s^d$ assigns probability $1$ to $d_s$.
Using the results of~\cite{EKVY08}, we prove the following lemma.
\begin{lem}\label{lem:sum-is-yC}
The existence of a strategy $\varrho$
satisfying 
$
\Ex\varrho{s_0}{\lrIf{\vec{r}}}
\geq
\vec{v}
$
implies the existence
of a (possibly randomized) memoryless strategy $\zeta$ for $G'$ such that
\begin{equation}\small
\label{eq:sum-is-yC}
\sum_{s\in C} \Pr\zeta{s_0}{\text{$\reach(d_s)$}} = y_C .
\end{equation}
\end{lem}
\begin{proof}
We give a proof by contradiction. Note that the proof structure is similar to the proof of direction 3$\Rightarrow$1 of Theorem 3.2 in \cite{EKVY08}.
Let $C_1,\ldots C_n$ be all MECs of $G$, and let $X\subseteq \Rset^n$ be the set of all vectors $(x_1,\ldots,x_n)$ for which
there is a strategy $\bar\sigma$ in $G'$ such that $\Pr{\bar\sigma}{s_{0}}{\bigcup_{s\in C_i}\reach(d_s)} \ge x_i$ for all $1\le i \le n$.
For a contradiction, suppose
$(y_{C_1},\ldots,y_{C_n}) \not\in X$. By \cite[Theorem 3.2]{EKVY08} the set $X$ can be described as a set of solutions of a linear program, and hence it is convex.
By the separating hyperplane theorem~\cite{Boyd04} there are weights $w_1,\ldots,w_n$ such that 
$\sum_{i=1}^n y_{C_i}\cdot w_i > \sum_{i=1}^n x_i\cdot w_i$ for every $(x_1,\ldots, x_n)\in X$.

We define a reward function $r$ by
$r(a)=w_i$ for an action $a$ from $C_i$, where $1\le i\le n$, and $r(a)=0$ for actions not in any MEC. Observe that the mean payoff of 
any run that eventually stays in a MEC $C_i$ is $w_i$, and so the expected mean payoff w.r.t. $r$ under $\varrho$ is 
$\sum_{i=1}^n y_{C_i}\cdot w_i$. Because memoryless deterministic strategies suffice for maximising the (single-objective) expected mean payoff, there is also a
memoryless deterministic strategy
$\hat\sigma$ for $G$ that yields expected mean payoff w.r.t. $r$ equal to $z\ge\sum_{i=1}^n y_{C_i}\cdot w_i$. We now define a strategy $\bar\sigma$ for $G'$
to mimic $\hat\sigma$ until a BSCC is reached, and when a BSCC is reached, say along a path $w$, the strategy $\bar\sigma$
takes the action $a^d_{\last(w)}$.
Let $x_i=\Pr{\bar\sigma}{s_{0}}{\bigcup_{s\in C_i}\reach(d_s)}$. Due to the construction of $\bar\sigma$ we have 
$x_i$ is equal to the probability of runs that eventually stay in $C_i$ under $\hat\sigma$:
this follows because once a BSCC is reached on a path $w$, every run $\omega$ extending $w$
has an infinite suffix containing only states from the MEC containing the state $\last(w)$.
Hence $\sum_{i=1}^n x_i\cdot w_i = z$. However, by the choice of the weights $w_i$ we get that $(x_1,\ldots,x_n)\not\in X$,
and hence a contradiction, because $\bar\sigma$ witnesses that $(x_1,\ldots,x_n)\in X$.

Hence, we have obtained that there is some (possibly memory-dependent) strategy $\zeta$, and using \cite[Theorem 3.2]{EKVY08} we get that there
also is a memoryless strategy $\zeta$ with the required properties. This completes the proof of Lemma~\ref{lem:sum-is-yC}.
\end{proof}

We now proceed with the proof of Proposition~\ref{prop:strat-sol}.
Let $U_a$ be a function over the runs in $G'$ returning the (possibly infinite) number of times
the action $a$ is used.
We are now ready to define the assignment for the 
variables $y_\chi$ of $L$.
\begin{align*}
y_a
&
\coloneqq
\Ex\zeta{s_0}{U_a}
&&
\text{for all $a\in A$}
\\
y_s
&
\coloneqq
\Ex\zeta{s_0}{U_{a^d_s}}
=
\Pr\zeta{s_0}{\text{$\reach(d_s)$}}
&&
\text{for all $s\in S$}
.
\end{align*}

Note that~\cite[Lemma~3.3]{EKVY08} ensures that all
$y_a$ and $y_s$ are indeed well-defined finite values, and satisfy
Equations~(\ref{eq:ya}) of~$L$.
Equations~(\ref{eq:yC}) of~$L$ are satisfied
due to Equations~(\ref{eq:sum-is-yC}) and~(\ref{eq:fa-yc}).
Equations~(\ref{eq:sum-is-yC}) together with
$\sum_{a\in A\cap C} f(a){=}1$ imply Equations~(\ref{eq:ys1}) of~$L$.
This completes the proof of Proposition~\ref{prop:strat-sol}. 
\end{proof}
\medskip

The item A.1 in Section~\ref{sec-exp-obj} follows directly from 
Theorem~\ref{thrm-exp}.
Let us analyze A.2. Suppose $\vec{v}$ is a point of the Pareto curve.
Consider the system $L'$ of linear inequalities obtained from $L$
by replacing constants $\vec{v}_i$ in Inequalities~(\ref{eq:rew}) 
with new variables $z_i$. Let $Q\subseteq\Rset^n$ be the projection 
of the set of solutions of $L'$ to $z_1,\ldots,z_n$. From
Theorem~\ref{thrm-exp}
and the definition of Pareto curve,
the (Euclidean) distance of $\vec{v}$ to $Q$ is~$0$.
Because the set of solutions of $L'$ is a closed set, $Q$ is also closed
and thus $\vec{v}\in Q$.
This gives us a solution to $L$ with variables $z_i$ having values $\vec{v}_i$,
and we can use Theorem~\ref{thrm-exp}
to get a strategy witnessing that $\vec{v}\in\AcEx(\lrIf{\vec{r}})$. 

Now consider the items A.3 and A.4.
The system $L$ is linear, and hence the problem whether 
$\vec{v} \in \AcEx(\lrIf{\vec{r}})$ is decidable in polynomial time
by employing polynomial-time algorithms for linear programming.
A 2-memory stochastic-update strategy $\sigma$ satisfying 
$\Ex{\sigma}{s}{\lrIf{\vec{r}}} \geq \vec{v}$ can be computed as follows
(note that the proof of Proposition~\ref{prop:sol-strat} is \emph{not} fully 
constructive, so we cannot apply this proposition immediately).
First, we find a solution of the system $L$, and we denote by 
$\bar{x}_a$ the value assigned to $x_a$. Let 
$(T_1,B_1),\ldots,(T_n,B_n)$ be the end components such 
that $a\in \bigcup_{i=1}^n B_i$
iff $\bar{x}_a>0$, and $T_1,\ldots,T_n$ are pairwise disjoint. 
We construct another system of linear inequalities consisting of
Equations~(1) of $L$ and the equations
$
 \sum_{s\in T_i} y_s = \sum_{s\in T_i}\sum_{a\in \act{s}} \bar{x}_a
$
for all $1\le i\le n$. Due to \cite{EKVY08}, there is a solution to
this system iff in the MDP $G'$ from the proof of
Proposition~\ref{prop:sol-strat} there is a strategy that for every
$i$ reaches $d_s$ for $s\in T_i$ with probability $\sum_{s\in
  T_i}\sum_{a\in \act{s}} \bar{x}_a$.  Such a strategy indeed exists
(consider, e.g., the strategy $\sigma'$ from the proof of
Proposition~\ref{prop:sol-strat}). Thus, there is a solution to the above
system and we can denote by $\hat{y}_s$ and $\hat{y}_a$ the values
assigned to $y_s$ and $y_a$. We define $\sigma$ by
\[\begin{array}{lcl}\small
 \sigma_n(s,m_1)(a)& = &\bar{y}_a / \sum_{a'\in \act{s}} \bar{y}_{a'}\\
 \sigma_n(s,m_2)(a)& = &\bar{x}_a / \sum_{a'\in \act{s}} \bar{x}_{a'} %
\end{array}\]
and further
$
 \sigma_u(a,s,m_1)(m_2){=}y_s,
 \sigma_u(a,s,m_2)(m_2){=}1,
$
and the initial memory distribution assigns $(1-y_{s_0})$ and $y_{s_0}$ to $m_1$ and $m_2$, respectively. Due to
\cite{EKVY08} we have 
\[
 \Pr\sigma{s_0}{\text{change memory to $m_2$ in $s$}}=\hat{y}_s,
\]
and the rest follows similarly as in the proof of 
Proposition~\ref{prop:sol-strat}.

The item A.5  can be proved as follows:
To test that $\vec{v}\in\AcEx(\lrIf{\vec{r}})$ lies in
the Pareto curve we turn the system $L$
into a linear program $LP$
by adding the objective to maximize
$
 \sum_{1\le i\le k}\sum_{a\in A} x_{a}\cdot\vec{r}_i(a)
.
$
Then we check that there is no better solution than
$\sum_{1\le i \le k} \vec{v}_i$.

Finally, the item A.6 is obtained by considering the system $L'$ above
and computing all exponentially many vertices of the polytope of all
solutions. Then we compute projections of these vertices onto the dimensions
$z_1,\ldots,z_n$ and retrieve all the maximal vertices. Moreover, if for every
$\vec{v} \in \{\ell \cdot \vare \mid \ell \in \Zset \wedge {-}M_r \le \ell\cdot\vare \le M_r\}^k$ where $M_r = \max_{a\in A}\max_{1\le i\le k} |\vec{r}_i(a)|$ we decide whether
$\vec{v}\in \AcEx(\lrIf{\vec{r}})$, we can easily construct 
an $\varepsilon$-approximate Pareto curve.

\subsection{Deterministic-update Strategies for Expectation Objectives}
\label{app-expsem-det-up}

We now show that for expectation objectives, finite-memory deterministic update
strategies suffice. This is captured in the following proposition.

\begin{prop}\label{prop:sol-strat-det}
Every nonnegative solution of the system $L$
induces a finite-memory deterministic-update
strategy $\sigma$ satisfying
$\Ex\sigma{s_0}{\lrIf{\vec{r}}}\geq \vec{v}$.
\end{prop}

\begin{proof}
The proof proceeds almost identically to the proof
of Proposition~\ref{prop:sol-strat}.
Let us recall the important steps from that proof first.
There we worked with the numbers $\bar{x}_a$, $a\in A$,
which, assigned to the variables $x_a$, formed a part of the solution to $L$.
We also worked with two important strategies.
The first one, a finite-memory deterministic-update strategy $\zeta$,
made sure that, starting in $s_0$, a run stays in a MEC $C$
forever with probability
$y_C=\sum_{a\in A\cap C}\bar{x}_a$.
The second one, a memoryless strategy $\sigma'$, had the
property that when the starting distribution
was $\alpha(s)\coloneqq\bar{x}_s= \sum_{a \in\act{s}}\bar{x}_a$
then
$\Ex{\sigma'}{\alpha}{\lrIf{\vec{r}}}\geq \vec{v}$.
\footnote{Here we extend the notation in a straightforward way
from a single initial state to a general initial distribution, $\alpha$.}
To produce the promised finite-memory deterministic-update
strategy $\sigma$ we now have to combine the strategies
$\zeta$ and $\sigma'$ using only deterministic memory updates.

We now define the strategy $\sigma$.
It works in three phases. First, it reaches every MEC $C$
and stays in it with the probability $y_C$.
Second, it prepares the distribution $\alpha$, and finally third,
it switches to $\sigma'$.
It is clear how the strategy is defined in the third phase.
As for the first phase, this is also identical to what we did
in the proof of Proposition~\ref{prop:sol-strat} for $\bar\sigma$:
The strategy $\sigma$
follows the strategy $\zeta$ from beginning until in the associated finite
state Markov chain $G^\zeta$ a bottom strongly connected component (BSCC)
is reached. At that point the run has already entered its final MEC $C$
to stay in it forever, which happens with probability $y_C$.

The last thing to solve is thus the second phase.
Two cases may occur. Either there is a state $s\in C$
such that $|\act{s}\cap C|>1$, i.e., there are at least two actions
the strategy can take from $s$ without leaving $C$.
Let us denote these actions $a$ and $b$.
Consider an enumeration
$
C
=
\{s_1,\ldots,s_k\}
$
of vertices of $C$.
Now we define the second phase of $\sigma$ when in $C$.
We start with defining the memory used in the second phase.
We symbolically represent the possible contents of the memory as
$\{\WAIT_1,\ldots,\WAIT_k,\SWITCH_1,\ldots,\SWITCH_k\}$.
The second phase then starts with the memory set to
$\WAIT_1$.
Generally, if the memory is set to $\WAIT_i$
then $\sigma$ aims at reaching
$s$ with probability $1$. This is possible (since $s$ is in the same MEC)
and it is a well known fact that it can be done without using memory.
On visiting $s$, the strategy chooses the action $a$ with probability
$
{\bar{x}_{s_i}}%
/
(y_C-\sum_{j=1}^{i-1}\bar{x}_{s_j})
$
and the action $b$ with the remaining probability.
In the next step the deterministic update function
sets the memory either to $\SWITCH_i$ or $\WAIT_{i+1}$, depending
on whether the last action seen is $a$ or $b$, respectively.
(Observe that if $i=k$ then the probability of taking
$b$ is $0$.)
The memory set to $\SWITCH_i$ means that the strategy aims
at reaching $s_i$ almost surely, and upon doing so, the strategy
switches to the third phase, following $\sigma'$.
It is easy to observe that on the condition of staying in $C$
the probability of switching to the third phase in some $s_i\in C$
is $\bar{x}_{s_i} / y_C$, thus the unconditioned probability
of doing so is $\bar{x}_{s_i}$, as desired.

The remaining case to solve is when
$|\act{s}\cap C|=1$ for all $s\in C$.
But then switching to the third phase is solved trivially
with the right probabilities, because staying in $C$ inevitably
already means mimicking $\sigma'$.
\end{proof}

\section{Solution for Satisfaction Objectives}
\label{sec-probsem}
In this section we prove the items B.1--B.6 of
Section~\ref{sec-results-sat}.  Let us fix an MDP $G$, a vector of
rewards, $\vec{r}=(r_1,\ldots,r_k)$, and an initial state
$s_0$.  We start by assuming that the MDP $G$ is strongly connected 
(i.e., $(S,A)$ is an end component).
\begin{prop}\label{prop:prob-ex}
  Assume that $G$ is strongly connected and that there is a strategy
  $\pi$ such that $\Pr{\pi}{s_0}{\lrIf{\vec{r}} \ge
    \vec{v}}\,>\,0$. Then the following is true.
\begin{enumerate}
\item[1.] There is a strategy $\xi$ satisfying $\Pr{\xi}{s}{\lrIf{\vec{r}}\ge\vec{v}}=1$ for all $s\in S$.
\item[2.] For each $\varepsilon{>}0$ there is a memoryless randomized strategy $\xi_{\varepsilon}$ that for all $s\in S$ satisfies $\Pr{\xi_{\varepsilon}}{s}{\lrIf{\vec{r}}\ge\vec{v} - \vec{\varepsilon}} =1$.
\end{enumerate}
Moreover, the problem whether there is some $\pi$ such that 
$\Pr{\pi}{s_0}{\lrIf{\vec{r}}\ge\vec{v}}>0$ is decidable in polynomial time.
Strategies $\xi_{\varepsilon}$ are computable in time polynomial in the size 
of $G$, the size of the binary representation of $\vec{r}$, and  
$\frac{1}{\varepsilon}$.
\end{prop}
\begin{proof}
By~\cite{Cha07b,GH10} we get that $\Pr{\pi}{s_0}{\lrIf{\vec{r}} \ge
  \vec{v}}\,>\,0$ implies that there is a strategy $\xi$ such that
$\Pr{\xi}{s_0}{\lrIf{\vec{r}} \ge \vec{v}}\,=\,1$:
Since $\lrIf{\vec{r}} \ge \vec{v}$ is a tail or prefix-independent 
function, it follows from the results of~\cite{Cha07b} that 
if $\Pr{\pi}{s_0}{\lrIf{\vec{r}} \ge \vec{v}}\,>\,0$, then there 
exists a state $s$ in the MDP with value~1, i.e., there exists $s$ such 
that $\sup_{\pi} \Pr{\pi}{s}{\lrIf{\vec{r}} \ge \vec{v}}=1$.
It follows from the results of~\cite{GH10} that in MDPs with tail functions,
optimal strategies exist and thus it follows that there exist 
a strategy $\pi_1$ from $s$ such that $\Pr{\pi_1}{s}{\lrIf{\vec{r}} \ge \vec{v}} =1$.
Since the MDP is strongly connected, the state $s$ can be reached 
with probability~1 from $s_0$ by a strategy $\pi_2$. 
Hence the strategy $\pi_2$, followed by the strategy $\pi_1$ after reaching 
$s$, is the witness strategy $\pi'$ such that 
$\Pr{\pi'}{s_0}{\lrIf{\vec{r}} \ge \vec{v}}\,=\,1$.

This gives us item 1. of
Proposition~\ref{prop:prob-ex} and also immediately implies
$\vec{v}\in \AcEx(\lrIf{\vec{r}})$.  It follows that there are
nonnegative values $\bar{x}_a$ for all $a\in A$ such that assigning
$\bar{x}_a$ to $x_a$ solves Equations~(\ref{eq:xa}) and~(\ref{eq:rew}) 
of the system $L$ (see Figure~\ref{system-L}).  Let us assume,
w.l.o.g., that $\sum_{a\in A} \bar{x}_a=1$.

Lemma~\ref{lem:xsol} gives us a memoryless randomized strategy $\zeta$
such that for all BSCCs $D$ of $G^{\zeta}$, all $s\in D\cap S$ and all
$a\in D\cap A$ we have that
$\freq{\zeta}{s}{a}=\frac{\bar{x}_a}{\sum_{a\in D\cap A} \bar{x}_a}$.
We denote by $\freq{\zeta}{D}{a}$ the value $\frac{\bar{x}_a}{\sum_{a\in
    D\cap A} \bar{x}_a}$.

Now we are ready to prove the item 2 of
Proposition~\ref{prop:prob-ex}.  Let us fix $\varepsilon>0$.  We
obtain $\xi_{\varepsilon}$ by a suitable perturbation of the strategy
$\zeta$ in such a way that all actions get positive probabilities and
the frequencies of actions change only slightly.
There exists an arbitrarily small
(strictly) positive solution $x'_a$ of Equations~(\ref{eq:xa}) of
the system $L$ (it suffices to consider a strategy $\tau$ which always
takes the uniform distribution over the actions
in every state and then assign $\freq{\tau}{s_0}{a}/N$ to
$x_a$ for sufficiently large $N$).
As the system of Equations~(\ref{eq:xa}) is linear and homogeneous,
assigning $\bar{x}_a+x'_a$ to $x_a$ also solves this system and
Lemma~\ref{lem:xsol} gives us a strategy $\xi_{\varepsilon}$
satisfying
$
\freq{\xi_{\varepsilon}}{s_0}{a}=(\bar{x}_a+x'_a)/X
$
where $X=\sum_{a'\in A}\bar{x}_{a'}+x'_{a'}=1+\sum_{a'\in
  A}x'_{a'}$. We may safely assume that $\sum_{a'\in A}
  x'_{a'}\leq \frac{\varepsilon}{2\cdot M_r}$
where $M_r=\max_{a\in A} \max_{1\leq i\leq k} |\vec{r}_i(a)|$. Thus,
we obtain
\begin{equation}\label{eq:freq-dom}
\sum_{a\in A} \freq{\xi_{\varepsilon}}{s_0}{a}\cdot \vec{r}_i(a) \geq \vec{v}_i - \varepsilon
\end{equation}
by the following sequence of (in)equalities.
{\small
\begin{align*}
\sum_{a\in A} & \freq{\xi_{\varepsilon}}{s_0}{a}\cdot \vec{r}_i(a) \\
& =  \sum_{a\in A} \frac{\bar{x}_a+x'_a}{X}\cdot \vec{r}_i(a) \tag{def}\\
& =   \frac{1}{X}\cdot \sum_{a\in A} \bar{x}_a\cdot \vec{r}_i(a)+\frac{1}{X}\cdot \sum_{a\in A}x'_a\cdot \vec{r}_i(a) \tag{rearranging}\\
  & =   \Big(\sum_{a\in A} \bar{x}_a\cdot \vec{r}_i(a) + \frac{1-X}{X}\cdot\sum_{a\in A} \bar{x}_a\cdot \vec{r}_i(a)\Big)
    +\frac{1}{X}\cdot \sum_{a\in A}x'_a\cdot \vec{r}_i(a) \tag{rearranging}\\
  & \geq   \sum_{a\in A} \bar{x}_a\cdot \vec{r}_i(a)
    -\Big|\frac{1-X}{X}\cdot\sum_{a\in A} \bar{x}_a\cdot \vec{r}_i(a)\Big|
    -\Big|\frac{1}{X}\cdot \sum_{a\in A}x'_a\cdot \vec{r}_i(a)\Big| \tag{property of abs. value}\\
  & \geq   \sum_{a\in A} \bar{x}_a\cdot \vec{r}_i(a)
    -\Big(\Big|(1-X)\cdot\sum_{a\in A} \bar{x}_a\cdot \vec{r}_i(a)\Big|+\Big|\sum_{a\in A}x'_a\cdot \vec{r}_i(a)\Big|\Big) \tag{from $X>1$}\\
  & \geq   \sum_{a\in A} \bar{x}_a\cdot \vec{r}_i(a)
    - \Big((1-X)\cdot \sum_{a\in A} \bar{x}_a\cdot |\vec{r}_i(a)|+\sum_{a\in A}x'_a\cdot |\vec{r}_i(a)|\Big) \tag{prop. of $|{\cdot}|$ and $X>1$}\\
  & \geq   \sum_{a\in A} \bar{x}_a\cdot \vec{r}_i(a)
    - \Big((1-X)\cdot M_r+\sum_{a\in A}x'_a\cdot M_r\Big) \tag{property of $M_r$}\\
  & \geq   \sum_{a\in A} \bar{x}_a\cdot \vec{r}_i(a)
   - \bigg(\Big(\sum_{a\in A}x'_a\Big)\cdot M_r+\Big(\sum_{a\in A}x'_a\Big)\cdot 
  \mathit{M_r}\bigg) \tag{property of $X$ and rearranging}\\
  & =  \sum_{a\in A} \bar{x}_a\cdot \vec{r}_i(a)- 2 \cdot \Big(\sum_{a\in A}x'_a\Big)\cdot M_r \tag{rearranging}\\
  & \geq \vec{v}_i - 2\cdot \Big(\sum_{a\in A}x'_a\Big)\cdot M_r \tag{property of $\vec{v}$}\\
  & \geq \vec{v}_i - \varepsilon \tag{property of $\vare$}
\end{align*} }

As $G^{\xi_\vare}$ is strongly connected, almost all runs $\omega$ of
$G^{\xi_\vare}$ initiated in $s_0$ satisfy
\[
\lrIf{\vec{r}}(\omega)\quad = \quad \sum_{a\in A} \freq{\xi_{\varepsilon}}{s_0}{a}\cdot \vec{r}(a)\quad \geq\quad \vec{v} - \vec{\varepsilon}
.
\]
This finishes the proof of item~2.

Concerning the complexity of computing $\xi_{\varepsilon}$, note that
the binary representation of every coefficient in~$L$ 
has only polynomial length. As
$\bar{x}_a$'s are obtained as a solution of (a part of) $L$, standard
results from linear programming imply that each $\bar{x}_a$ has a
binary representation computable in polynomial time. The numbers
$x'_a$ are also obtained by solving a part of~$L$ and restricted by
$\left|\sum_{a'\in A} x'_{a'}\right|\leq \frac{\varepsilon}{2\cdot
M_r}$ which allows to compute a binary representation of $x'_a$ in
polynomial time.  The strategy $\xi_{\varepsilon}$, defined in the
proof of Proposition~\ref{prop:prob-ex}, assigns to each action only small
arithmetic expressions over $\bar{x}_a$ and $x'_a$. Hence, $\xi_{\varepsilon}$
is computable in polynomial time.

To prove that the problem whether there is some $\xi$ such that
$\Pr{\xi}{s_0}{\lrIf{\vec{r}}\ge\vec{v}}>0$ is decidable in polynomial
time, we show that whenever $\vec{v}\in\AcEx(\lrIf{\vec{r}})$, then
$(1,\vec{v})\in \AcPr(\lrIf{\vec{r}})$. This gives us a
polynomial-time algorithm by applying Theorem~\ref{thrm-exp}.  Let
$\vec{v}\in\AcEx(\lrIf{\vec{r}})$. We show that there is a strategy
$\xi$ such that $\Pr{\xi}{s}{\lrIf{\vec{r}} \geq\vec{v}} = 1$.

Since $\vec{v}\in\AcEx(\lrIf{\vec{r}})$, there are nonnegative
rational values $\bar{x}_a$ for all $a\in A$ such that assigning
$\bar{x}_a$ to $x_a$ solves Equations~(\ref{eq:xa}) and~(\ref{eq:rew}) 
of the system $L$.  Assume, 
without loss of generality, that $\sum_{a\in A} \bar{x}_a=1$.

Given $a\in A$, let $I_a:A\rightarrow\{0,1\}$ be a function given by
$I_a(a)=1$ and $I_a(b)=0$ for all $b\neq a$.  For every $i\in \Nset$,
we denote by $\xi_i$ a memoryless randomized strategy satisfying
$\Pr{\xi_i}{s}{\lrIf{I_a} \ge \bar{x}_a - 2^{-i-1}}=1$.  Note that for
every $i\in \Nset$ there is $\stabi{i}\in\Nset$ such that for all
$a\in A$ and $s\in S$ we get
\[
 \Pr{\xi_{i}}{s}{\inf_{T \ge \stabi{i}} \frac{1}{T}\sum_{t=0}^T I_a(A_t) \ge \bar{x}_a - 2^{-i}} \ge 1-2^{-i}
 .
\]
Now let us consider a sequence $n_0,n_1,\ldots$ of numbers where
$n_i\ge \kappa_i$ and $\frac{\sum_{j<i} n_j}{n_i}\leq 2^{-i}$ and
$\frac{\stabi{i+1}}{n_i}\leq 2^{-i}$.  We define $\xi$ to
behave as $\xi_1$ for the first $n_1$ steps, then as $\xi_2$ for the
next $n_2$ steps, then as $\xi_3$ for the next $n_3$ steps, etc. In
general, denoting by $N_i$ the sum $\sum_{j<i} n_j$, the strategy
$\xi$ behaves as $\xi_i$ between the $N_i$'th step (inclusive) and
$N_{i+1}$'th step (non-inclusive).

Let us give some intuition behind $\xi$. The numbers in the sequence $n_0,n_1,\ldots$ grow rapidly
so that after $\xi_i$ is simulated for $n_i$ steps, the part of the history when $\xi_j$ for $j<i$ were simulated
becomes relatively small and has only minor impact on the current average reward (this is
ensured by the condition $\frac{\sum_{j<i} n_j}{n_i}\leq 2^{-i}$). This gives us that almost every
run has infinitely many prefixes on which the average reward w.r.t. $I_a$ is arbitrarily close to
$\bar{x}_a$ infinitely often. To get that $\bar{x}_a$ is also the limit-average reward,
one only needs to be careful when the strategy $\xi$ ends behaving
as $\xi_i$ and starts behaving as $\xi_{i+1}$, because then up to the $\kappa_{i+1}$ steps we have
no guarantee that the average reward is close to $\bar{x}_a$. This part is taken
care of by picking $n_i$ so large that the contribution (to the average reward) of the $n_i$ steps according to
$\xi_i$ prevails over fluctuations introduced by the first
$\kappa_{i+1}$ steps according to $\xi_{i+1}$ (this is ensured by the condition 
$\frac{\stabi{i+1}}{n_i}\leq 2^{-i}$).

Let us now prove the correctness of the definition of $\xi$ formally.
We prove that almost all runs $\omega$ of $G^{\xi}$ satisfy 
\[
\liminf_{T\rightarrow\infty} \frac{1}{T}\sum_{t=0}^T I_a(A_t(\omega))\geq \bar{x}_a
.
\]
Denote by $E_i$ the set of all runs $\omega=s_0a_0s_1a_1\ldots$ of $G^{\xi}$ such that for some $\stabi{i}\le d \le n_i$ we have
\[
 \frac{1}{d}\sum_{j=N_i}^{N_i+d}I_a(a_{j})\quad < \quad \bar{x}_a - 2^{-i}
 .
\]
We have $\Pr{\xi}{s_0}{E_i} \le 2^{-i}$ and thus
$\sum_{i=1}^{\infty} \Pr{\xi}{s_0}{E_i}=\frac{1}{2}<\infty$. By Borel-Cantelli lemma~\cite{Royden88}, almost surely only
finitely many of $E_{i}$ take place. Thus, almost
every run $\omega=s_0a_0s_1a_1\ldots$ of $G^{\xi}$ satisfies the following: there is $\ell$ such that for all $i\ge\ell$ and all $\stabi{i}\le d \le n_i$
we have that
\[
 \frac{1}{d}\sum_{j=N_i}^{N_i+d}I_a(a_{j})\quad \geq \quad \bar{x}_a - 2^{-i}
 .
 \]
Consider $T\in \Nset$ such that $N_{i}\leq T<N_{i+1}$ where $i>\ell$. 
We need the following inequality
\begin{equation}\label{eq:fin}
\frac{1}{T} \sum_{t=0}^T  I_a(a_t)\quad \geq\quad (\bar{x}_a-2^{-i})(1-2^{1-i})
\end{equation}
which can be proved as follows.
First, note that
\[
\frac{1}{T} \sum_{t=0}^T  I_a(a_t) \quad \geq\quad
\frac{1}{T}\sum_{t=N_{i-1}}^{N_{i}-1} I_a(a_t)+\frac{1}{T}\sum_{t=N_i}^{T} I_a(a_t)
\]
and that
\begin{align*}
\frac{1}{T}  \sum_{t=N_{i-1}}^{N_{i}-1}  I_a(a_t) & = \frac{1}{n_i} \sum_{t=N_{i-1}}^{N_{i}-1} I_a(a_t)\cdot \frac{n_i}{T} 
   \geq (\bar{x}_a-2^{-i}) \frac{n_i}{T}
\end{align*}
which gives
\begin{equation}\label{eq:main}
\frac{1}{T} \sum_{t=0}^T  I_a(a_t) \ \geq \  
(\bar{x}_a-2^{-i}) \frac{n_i}{T}\, +\, \frac{1}{T}\sum_{t=N_i}^{T} I_a(a_t)
.
\end{equation}
Now, we distinguish two cases.
First, if $T-N_i\leq \stabi{i+1}$, then 
\begin{align*}
 \frac{n_i}{T}\geq\frac{n_i}{N_{i-1}+n_i +\stabi{i+1}} = 1-\frac{N_{i-1}+\stabi{i+1}}{N_{i-1}+n_i+\stabi{i+1}} \geq (1-2^{1-i})
\end{align*}
and thus, by Equation~(\ref{eq:main}),
\[
\frac{1}{T} \sum_{t=0}^T  I_a(a_t) \quad \geq\quad (\bar{x}_a-2^{-i})(1-2^{1-i})
.
\]
Second, if $T-N_i\geq \stabi{i+1}$, then 
\begin{align*}
\frac{1}{T}\sum_{t=N_i+1}^{T} I_a(a_t) & =\frac{1}{T-N_i}\sum_{t=N_i+1}^{T} I_a(a_t)\cdot \frac{T-N_i}{T} \\
 &  \geq (\bar{x}_a-2^{-i-1})\left(1-\frac{N_{i-1}+n_i}{T}\right) \\
 &  \geq (\bar{x}_a-2^{-i-1})\left(1-2^{-i}-\frac{n_i}{T}\right)
\end{align*}
and thus, by Equation~(\ref{eq:main}),
\begin{align*}
\frac{1}{T} \sum_{t=0}^T  I_a(a_t) & \geq 
(\bar{x}_a-2^{-i}) \frac{n_i}{T} + (\bar{x}_a-2^{-i-1})\left(1-2^{-i}-\frac{n_i}{T}\right) \\
& \geq (\bar{x}_a-2^{-i}) \left(\frac{n_i}{T} + \left(1-2^{-i}-\frac{n_i}{T}\right) \right) \\
& \geq (\bar{x}_a-2^{-i}) (1-2^{-i}) 
\end{align*}
which finishes the proof of Equation~(\ref{eq:fin}).

Since the sum in Equation~(\ref{eq:fin}) converges to $\bar{x}_a$ as $i$ (and thus also $T$) 
goes to $\infty$, we obtain
\[
 \liminf_{T\rightarrow\infty} \frac{1}{T}\sum_{t=0}^{T}I_a(a_t) \ge \bar{x}_a
 .
\]
\vspace{-8mm}

\end{proof}
\smallskip

\begin{figure}
 \begin{center}
  \begin{tikzpicture}
 \node[state] at (0,0) (l) {$s_1$};
 \node[state] at (3,0) (r) {$s_2$};
 \node[distr] at (1.5,0.4) (t) {};
 \node[distr] at (1.5,-0.4) (b) {};

 \node[distr] at (-0.7,0) (ll) {};
 \node[distr] at (3.7,0) (rr) {};
 
 \draw[trarr] (l) |- (t) node[pos=0.8,below] {$a_1$} -| (r);
 \draw[trarr] (r) |- (b) node[pos=0.8,above] {$a_2$} -| (l);

 \draw[trarr] (l) -- +(-0.7,-0.4) -- (ll) node[midway,left] {$b_1$} -- +(0,0.4) -- (l);
 \draw[trarr] (r) -- +(0.7,-0.4) -- (rr) node[midway,right] {$b_2$} -- +(0,0.4) -- (r);
\end{tikzpicture}
 \end{center}
 \caption{MDP showing the need of infinite memory.\label{fig:infinite}}
\end{figure}
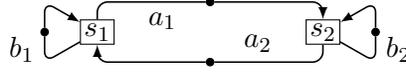

The strategy $\xi$ from the proof of Proposition~\ref{prop:prob-ex}
required infinite memory. We show that this may indeed be necessary,
i.e. it can be the case that $(\nu,\vec{v})\in \AcPr(\lrIf{\vec{r}})$
although there is no finite-memory strategy $\sigma$ satisfying
$\Pr{\sigma}{s}{\lrIf{\vec{r}} \geq \vec{v}} > \nu$ (and in fact not even finite-memory strategy satisfying $\Pr{\sigma}{s}{\lrIf{\vec{r}} \geq \vec{v}} > 0$).
\label{lab:infinite}
Consider the MDP from Figure~\ref{fig:infinite},
where the reward function $r_i$ (for $i\in\{1,2\}$)
returns $1$ for $b_i$ and $0$ for all other actions.
Let $s_1$ be the initial vertex.
It is easy to see that $(0.5,0.5)\in \AcEx(\lrIf{\vec{r}})$: consider for example a strategy that first
chooses both available actions in $s_1$ with uniform probabilities, and in subsequent steps chooses self-loops
on $s_1$ or $s_2$ deterministically. From the results presented above we subsequently get that
$(1,0.5,0.5)\in \AcPr(\lrIf{\vec{r}})$.

On the other hand, let $\sigma$ be arbitrary finite-memory strategy. The Markov chain it induces is by definition
finite and for each of its BSCC $C$ we have the following. One of the following then takes place:
\begin{itemize}
 \item $C$ contains both $s_1$ and $s_2$. Then by Ergodic theorem for almost every run $\omega$ we have
  $\lrIf{I_{a_1}}(\omega) + \lrIf{I_{a_2}}(\omega) > 0$, which means that
  $\lrIf{I_{b_1}}(\omega) + \lrIf{I_{b_2}}(\omega) < 1$, and thus necessarily
  $\lrIf{\vec{r}}(\omega)\not\ge(0.5,0.5)$.
 \item $C$ contains only the state $s_1$ (resp. $s_2$), in which case all runs that
  enter it satisfy $\lrIf{\vec{r}}(\omega)=(1,0)$ (resp. $\lrIf{\vec{r}}(\omega)=(0,1)$).
\end{itemize}
From the basic results of the theory of Markov chains we get $\Pr{\sigma}{s_1}{\lrIf{\vec{r}} \geq (0.5,0.5)} = 0$.

It is also easy to prove that $\varepsilon$-optimal strategies are not necessarily memoryless pure,
as the following lemma shows.
\begin{lem}
\label{lem-eps-mp}
There is an MDP $G$ a vector of reward functions $\vec{r}=(r_1,r_2)$, a number $\varepsilon>0$
and a vector $(\nu,\vec{v})\in \AcPr(\lrIf{\vec{r}})$
such that there is no memoryless-pure strategy $\sigma$ satisfying
$\Pr{\sigma}{s}{\lrIf{\vec{r}} \geq \vec{v} - \vec{\varepsilon}} > \nu - \vec{\varepsilon}$.
\end{lem}
\begin{proof}
  We can reuse $G$ and $\vec{r}$ showing the need of infinite memory for optimal strategies.
 We let $\nu=1$ and
$\vec{v}=(0.5,0.5)$. We have shown that
$(\nu,\vec{v})\in \AcPr(\lrIf{\vec{r}})$. Taking e.g. $\varepsilon=0.1$,
it is a trivial observation that no memoryless
pure strategy satisfies 
$\Pr{\sigma}{s}{\lrIf{\vec{r}} \geq  \vec{v} - \vec{\varepsilon}} > \nu - \vec{\varepsilon}$.
\end{proof}

We are now ready to prove the items B.1, B.3 and B.4.  Let
$C_1,\ldots,C_{\ell}$ be all MECs of $G$. We say
that a MEC $C_i$ is {\em good for $\vec{v}$} if there is a state $s$
of $C_i$ and a strategy $\pi$ satisfying $\Pr{\pi}{s}{\lrIf{\vec{r}}
  \ge \vec{v}}\,>\,0$ that never leaves $C_i$ when starting in $s$.
Using Proposition~\ref{prop:prob-ex}, we can decide in polynomial time
whether a given MEC is good for a given $\vec{v}$.  Let $\MECuni$ be
the union of all MECs good for $\vec{v}$.  Then, by
Proposition~\ref{prop:prob-ex}, there is a strategy $\xi$ such that
for all $s\in \MECuni$ we have $\Pr{\xi}{s}{\lrIf{\vec{r}} \ge
  \vec{v}}\,=\,1$ and for each $\varepsilon>0$ there is a memoryless
randomized strategy $\xi_{\varepsilon}$, computable in polynomial
time, such that for all $s\in \MECuni$ we have
$\Pr{\xi_{\varepsilon}}{s_0}{\lrIf{\vec{r}}\ge\vec{v} -
  \vec{\varepsilon}}=1$.

Consider a strategy $\tau$, computable in polynomial time, which maximizes the probability of reaching $\MECuni$.
Denote by $\sigma$ a strategy which behaves as $\tau$ before reaching $\MECuni$ and as $\xi$ afterwards.
Similarly, denote by $\sigma_{\varepsilon}$ a strategy 
which behaves as $\tau$ before reaching $\MECuni$ and as $\xi_{\varepsilon}$ afterwards. Note that
$\sigma_{\varepsilon}$ is computable in polynomial time.

Clearly, $(\nu,\vec{v})\in \AcPr(\lrIf{\vec{r}})$ iff $\Pr{\tau}{s_0}{\reach(\MECuni)}\geq \nu$ because $\sigma$ achieves $\vec{v}$
with probability $\Pr{\tau}{s_0}{\reach(\MECuni)}$. Thus, we obtain that
$
\nu \leq \Pr{\tau}{s_0}{\reach(\MECuni)} \leq 
\Pr{\xi_{\varepsilon}}{s_0}{\lrIf{\vec{r}}\ge\vec{v} - \vec{\varepsilon}}
$.

Finally, in order to decide whether $(\nu,\vec{v})\in \AcPr(\lrIf{\vec{r}})$, it suffices to decide whether
$\Pr{\tau}{s_0}{\reach(\MECuni)}\geq \nu$ in polynomial time.

Now we prove  item B.2.
Suppose $(\nu,\vec{v})$ is a vector of the Pareto curve. 
We let $\MECuni$ be the union of all MECs good for
$\vec{v}$. Recall that the Pareto curve constructed for expectation 
objectives is achievable (item A.2). Due to the correspondence between 
$\AcPr$ and $\AcEx$ in strongly connected MDPs we obtain the following. 
There is $\lambda>0$ such that for every MEC $D$ not contained in $\MECuni$,
every $s \in D$,
and every strategy $\sigma$ that does not leave $D$, it is possible to have
$\Pr{\sigma}{s}{\lrIf{\vec{r}} \geq\vec{u}} > 0$ only if
there is $i$ such that $\vec{v}_i-\vec{u}_i\ge\lambda$,
i.e., when $\vec{v}$ is greater than $\vec{u}$ by $\lambda$ in some component.
Thus, for every $\vare<\lambda$ and every strategy $\sigma$ such that
$\Pr{\sigma}{s_0}{\lrIf{\vec{r}} \geq\vec{v} -\vec{\vare}} \ge \nu-\vare$ 
it must be the case that
$\Pr{\sigma}{s_0}{\reach(\MECuni)}\ge\nu - \vare$. Because for 
single objective reachability
the optimal strategies exist, we get that there is a strategy $\tau$ satisfying
$\Pr{\tau}{s_0}{\reach(\MECuni)}\ge \nu$, and by using methods similar to the ones of the previous
paragraphs we obtain $(\nu,\vec{v})\in \AcPr(\lrIf{\vec{r}})$.

The polynomial-time algorithm mentioned in item B.5 works as follows.
First check whether $(\nu,\vec{v}) \in \AcPr(\lrIf{\vec{r}})$ and if not, return ``no''.
Otherwise, find all MECs good for $\vec{v}$ and compute the maximal probability of reaching
them from the initial state. If the probability is \emph{strictly} greater than
$\nu$, return ``no''. Otherwise, continue
by performing the following procedure for every $1\le i \le k$, where 
$k$ is the dimension of $\vec{v}$:
Find all MECs $C$ for which there is $\vare>0$ such that $C$ is good for $\vec{u}$, where $\vec{u}$ is obtained from
$\vec{v}$ by increasing the $i$-th component by $\vare$ (this can be done in polynomial time using linear programming).
Compute the maximal probability of reaching these MECs.
If for any $i$ the probability is \emph{at least} $\nu$, return ``no'', otherwise return ``yes''.

The first claim of B.6 follows from Running example (II).
We prove that the set
$
N
\coloneqq
\{ \nu \mid (\nu,\vec{v}) \in P \}
,
$
where $P$ is the Pareto curve for 
$\AcPr(\lrIf{\vec{r}})$,
is indeed finite.
As we already showed, for every fixed $\vec{v}$ there is a
union $\MECuni$ of MECs good for $\vec{v}$, and
$(\nu,\vec{v})\in\AcPr(\lrIf{\vec{r}})$
iff
the $\MECuni$ can be reached with probability at least $\nu$.
Hence $|N| \leq 2^{|G|}$, because the latter
is an upper bound on a number of unions of MECs in $G$.

To prove the other claims, let
$N$ be the set $\{ \nu \mid (\nu,\vec{v}) \in P \}$
where $P$ is the Pareto curve for 
$\AcPr(\lrIf{\vec{r}})$.

Let us consider a fixed $\nu \in N$.
This gives us a collection $R(\nu)$
of all unions $\MECuni$ of MECs which can be reached with
probability at least $\nu$.
For a MEC $\MEC$ let $Sol(\MEC)$ be the set $\AcEx(\lrIf{\vec{r}})$
of the MDP given by restricting $G$ to $\MEC$.
Further, for every $\MECuni \in R(\nu)$ we set
$
Sol(\MECuni)
\coloneqq
\bigcap_{\MEC \in \MECuni} Sol(\MEC)
.
$
Finally,
$
Sol(R(\nu))
\coloneqq
\bigcup_{\MECuni\in R(\nu)} Sol(\MECuni)
.
$
From the analysis above we already know that
$
Sol(R(\nu))
=
\{\vec{v} \mid (\nu,\vec{v}) \in \AcPr(\lrIf{\vec{r}}\}
.
$
As a consequence, $(\nu,\vec{v})\in P$
iff $\nu\in N$
and
$\vec{v}$ is maximal in $Sol(R(\nu))$
and $\vec{v}\notin Sol(R(\nu'))$ for any
$\nu' \in N,\ \nu'>\nu$.
In other words, $P$ is also the Pareto curve
of the set
$
Q
\coloneqq
\{ (\nu,\vec{v}) \mid \nu \in N, \vec{v}\in Sol(R(\nu))\}
.
$
Observe that $Q$ is a finite union of downward closures of bounded convex polytopes,
because every $Sol(\MECuni)$ is a bounded convex polytope.
Finally, observe that $N$ can be computed using the algorithms
for optimizing single-objective reachability.
Further, the inequalities defining $Sol(\MECuni)$
can also be computed using our results on $\AcEx$.
By a \emph{generalised convex polytope}
we denote a set of points described by a finite conjunction
of linear inequalities, which may be both strict and
non-strict.
\begin{clm}
\label{cl:facets}
Let $X$ be a generalised convex polytope.
The smallest convex polytope containing $X$
is its closure, $cl(X)$.
Moreover, the set $cl(X)\setminus X$ is a union
of some of the facets of $cl(X)$.
\end{clm}
\begin{proof}
Let $I$ by the set of inequalities defining $X$,
and denote by $I'$ the modification of this set where
all the inequalities are transformed to non-strict ones.
The closure $cl(X)$ indeed is a convex polytope, as it is described
by $I'$.
Since every convex polytope is closed, if it contains $X$
then it must contain also its closure. Thus $cl(X)$ is the smallest one
containing $X$.
Let $\alpha < \beta$ be a strict inequality from $I$.
By $I'(\alpha=\beta)$ we denote the set $I' \cup \{\alpha=\beta\}$.
The points of $cl(X)\setminus X$ form a union of convex polytopes,
each one given by the set $I'(\alpha=\beta)$ for some $\alpha<\beta \in I$.
Thus, it is a union of facets of $cl(X)$.
\end{proof}
The following lemma now finishes the proof of B.6:
\begin{lem}
\label{lem:pareto-GCP}
Let $Q$ be a finite union of bounded convex polytopes, $Q_1,\ldots,Q_m$.
Then its Pareto curve $P$ is
a finite union of bounded generalised convex polytopes, $P_1,\ldots,P_n$.
Moreover, if the inequalities describing
$Q_i$ are given,
then the inequalities describing $P_i$
can be computed.
\end{lem}
\begin{proof}
We proceed by induction on the number $m$ of components of $Q$.
If $m=0$ then $P=\emptyset$ is clearly a bounded convex polytope
easily described by arbitrary two incompatible inequalities.
For $m\geq1$ we denote set $Q'\coloneqq \bigcup_{i=1}^{m-1}Q_i$.
By the induction hypothesis, the Pareto curve of $Q'$
is some $P'\coloneqq \bigcup_{i=1}^{n'}P_i$
where every $P_i$, $1\leq i \leq n'$ is a bounded generalised
convex polytope, described by some set of linear inequalities.
Denote by $dom(X)$ the (downward closed) set of all points dominated
by some point of $X$.
Observe that $P$, the Pareto curve of $Q$, is the union of all points
which either are maximal in $Q_m$ and do not belong to $dom(P')$
(observe that $dom(P')=dom(Q')$), or are in $P'$ and do not belong
to $dom(Q_m)$.
In symbols:
\[
P
=
(\text{maximal from $Q_m$} \setminus dom(P'))
\cup
(P' \setminus dom(Q_m))
.
\]
The set $dom(P')$ of all $\vec{x}$ for which there is some $\vec{y}\in P'$
such that $\vec{y}\geq\vec{x}$ is a union of projections of generalised
convex polytopes -- just add the inequalities from the definition
of each $P_i$ instantiated with $\vec{y}$ to the inequality $\vec{y}\geq\vec{x}$,
and remove $\vec{x}$ by projecting.
Thus, $dom(P')$ is a union of generalised convex polytopes itself.
A difference of two generalised convex polytopes
is a union of generalised convex polytopes.
Thus the set ``$\text{maximal from $Q_m$} \setminus dom(P')$''
is a union of generalised bounded convex polytopes,
and for the same reasons so is $P' \setminus dom(Q_m)$.

Finally, let us show how to compute $P$.
This amounts to computing the projection, and the set difference.
For convex polytopes, efficient computing of projections
is a problem studied since the 19th century.
One of possible approaches, non-optimal from the complexity point of view,
but easy to explain, is by traversing the vertices of the
convex polytope and projecting them individually, and then
taking the convex hull of those vertices.
To compute a projection of a generalised convex polytope $X$,
we first take its closure $cl(X)$, and project the closure.
Then we traverse all the facets of the projection
and mark every facet to which at least one point of $X$ projected.
This can be verified by testing whether the inequalities defining the facet
in conjunction with the inequalities defining $X$ have a solution.
Finally, we remove from the projection all facets which
are not marked.
Due to Claim~\ref{cl:facets}, the difference of the projection
of $cl(X)$ and the projection of $X$ is a union of facets.
Every facet from the difference has the property that no point
from $X$ is projected to it. Thus we obtained the projection
of $X$.

Computing the set difference of two bounded generalised convex polytopes
is easier: Consider we have two polytopes, given by sets $I_1$ and $I_2$
of inequalities.
Then subtracting the second generalised convex polytope from the first is
the union of generalised polytopes given by the inequalities
$I_1 \cup \{\alpha \nprec \beta\}$, where $\alpha \prec \beta$ ranges over all
inequalities (strict or non-strict) in $I_2$.
\end{proof}

\section{A Note on Equivalence of Definitions of Strategies}
\label{app-strat-eq}
In this section we argue that the definitions of strategies
as functions $(SA)^*S \to \dist(A)$ and as triples 
$(\sigma_u,\sigma_n,\alpha)$ are interchangeable. 

Note that formally a strategy $\pi : (SA)^*S \to \dist(A)$
gives rise to a Markov chain $G^\pi$ with states
$(SA)^*S$ and transitions $w \tran{\sigma(w)(a)\cdot \delta(a)(s)} was$
for all $w\in (SA)^*S$, $a\in A$ and $s\in S$.
Given $\sigma=(\sigma_u,\sigma_n,\alpha)$ and 
a run $w=(s_0,m_0,a_0)(s_1,m_1,a_1)\ldots$ of $G^\sigma$ denote
$w[i]=s_0a_0s_1a_1\ldots s_{i-1}a_{i-1}s_i$. We define
$f(w) = w[0]w[1]w[2]\ldots$.

We need to show that for every strategy $\sigma=(\sigma_u,\sigma_n,\alpha)$
there is a strategy $\pi: (SA)^*S \to \dist(A)$ (and vice versa)
such that for every set of runs $W$ of $G^\pi$
we have $\Pr{\sigma}{s_0}{f^{-1}(W)}=\Pr{\pi}{s_0}{W}$. We only present
the construction of strategies and basic arguments, the technical
part of the proof is straightforward.

Given $\pi : (SA)^*S \to \dist(A)$, one can easily define 
a \emph{deterministic-update} strategy 
$\sigma = (\sigma_u,\sigma_n,\alpha)$ which uses 
memory $(SA)^*S$. The initial memory element is the initial state $s_0$,
the next move function is defined by $\sigma(s,w)=\pi(w)$, and the memory update
function $\sigma_u$ is defined by $\sigma_u(a,s,w)=was$. Reader can observe that
there is a naturally defined bijection between runs in $G^\pi$ and in $G^\sigma$, and that
this bijection preserves probabilities of sets of runs.

In the opposite direction, given $\sigma = (\sigma_u,\sigma_n,\alpha)$,
we define $\pi : (SA)^*S \to \dist(A)$ as follows. Given
$w=s_0a_0\ldots s_{n-1}a_{n-1}s_n\in (SA)^*S$ and $a\in A$, we denote by $U^w_a$ the set
of all paths in $G^\sigma$ that have the form
\[
 (s_0,m_0,a_0)(s_1,m_1,a_1)\ldots(s_{n-1},m_{n-1},a_{n_1})(s_n,m_n,a)
\]
for some $m_1,\ldots m_n$. We put $\pi(w)(a)=\frac{\Pr{\sigma}{s_0}{U^w_a}}{\sum_{a'\in A} \Pr{\sigma}{s_0}{U^w_{a'}}}$.
The key observation for the proof of correctness of this construction is that the probability
of $U^w_a$ in $G^\sigma$ is equal to probability of taking a path $w$ and then an action $a$ in $G^\pi$.

\section{Conclusions}
In this paper we have studied the problem of determining whether for a given MDP there exists a strategy achieving
a certain value in each of multiple given limit-average objective functions. We have concentrated on two different interpretations
of the functions, namely the expectation objectives and satisfaction objectives, and provided algorithms solving the problem.

The next step in this line of research is to implement and evaluate the algorithms. On the theoretical side, one could further
study the problem of existence of a strategy that simultaneously satisfies several expectation objective and satisfaction objectives,
or even combine the limit-average functions with different kinds of functions, such as $\omega$-regular objectives or cumulative reward objectives.

\medskip
\noindent
{\bf Acknowledgements.} The authors thank David Parker and Dominik 
Wojtczak for initial discussions on the topic.
T.~Br\'{a}zdil is supported by the Czech Science Foundation, grant
No~P202/12/P612.
K.~Chatterjee is supported by the 
Austrian Science Fund (FWF) Grant No P 23499-N23;
FWF NFN Grant No S11407-N23 (RiSE);
ERC Start grant (279307: Graph Games);
Microsoft faculty fellows award. V. Forejt is supported by a Royal Society Newton Fellowship and EPSRC project~EP/J012564/1.

\bibliographystyle{abbrv}
\bibliography{bib}
\end{document}